\documentclass[twocolumn,twocolappendix]{aastex631}

\pdfoutput=1

\usepackage{enumerate}
\usepackage{enumitem}
\usepackage{natbib,ifthen}
\usepackage{graphicx}
\usepackage{fancyhdr}
\usepackage{amssymb,amsmath}
\usepackage{multirow,bigdelim}
\usepackage{wasysym}  
\usepackage{amssymb}
\usepackage[T1]{fontenc}

\shorttitle{HDO Detection in the LMC}
\shortauthors{Sewi{\l}o et al.}

\begin{document}

\title{The Detection of Deuterated Water in the Large Magellanic Cloud with ALMA}

\correspondingauthor{Marta Sewi{\l}o}
\email{marta.m.sewilo@nasa.gov}

\author[0000-0003-2248-6032]{Marta Sewi{\l}o}
\affiliation{Exoplanets and Stellar Astrophysics Laboratory, NASA Goddard Space Flight Center, Greenbelt, MD 20771, USA}
\affiliation{Department of Astronomy, University of Maryland, College Park, MD 20742, USA}
\affiliation{Center for Research and Exploration in Space Science and Technology, NASA Goddard Space Flight Center, Greenbelt, MD 20771} 

\author[0000-0001-8913-925X]{Agata Karska}
\affiliation{Institute of Astronomy, Faculty of Physics, Astronomy and Informatics, Nicolaus Copernicus University, ul. Grudzi\k{a}dzka 5, 87-100 Toru\'{n}, Poland}

\author[0000-0003-1159-3721]{Lars E. Kristensen}
\affiliation{Niels Bohr Institute, Centre for Star \& Planet Formation, University of Copenhagen, {\O}ster Voldgade 5-7, 1350 Copenhagen K, Denmark}

\author[0000-0001-6752-5109]{Steven B. Charnley}
\affiliation{Astrochemistry Laboratory, NASA Goddard Space Flight Center, Greenbelt, MD 20771, USA}

\author[0000-0002-3925-9365]{C.-H. Rosie Chen}
\affiliation{Max-Planck-Institut f{\"u}r Radioastronomie, Auf dem H{\"u}gel 69 D-53121 Bonn, Germany}

\author[0000-0002-0861-7094]{Joana M. Oliveira}
\affiliation{Lennard-Jones Laboratories, Keele University, ST5 5BG, UK}

\author[0000-0001-8233-2436]{Martin Cordiner}
\affiliation{Astrochemistry Laboratory, NASA Goddard Space Flight Center, Greenbelt, MD 20771, USA}
\affiliation{Institute for Astrophysics and Computational Sciences, The Catholic University of America, Washington, DC 20064, USA}

\author[0000-0002-1143-6710]{Jennifer Wiseman}
\affiliation{Exoplanets and Stellar Astrophysics Laboratory, NASA Goddard Space Flight Center, Greenbelt, MD 20771, USA}

\author[0000-0002-3078-9482]{\'{A}lvaro S\'{a}nchez-Monge}
\affiliation{I. Physikalisches Institut der Universit{\"a}t zu K{\"o}ln, Z{\"u}lpicher Str. 77, 50937, K{\"o}ln, Germany}

\author[0000-0002-1272-3017]{Jacco Th. van Loon}
\affiliation{Lennard-Jones Laboratories, Keele University, ST5 5BG, UK}

\author[0000-0002-4663-6827]{Remy Indebetouw}
\affiliation{Department of Astronomy, University of Virginia, PO Box 400325, Charlottesville, VA 22904, USA}
\affiliation{National Radio Astronomy Observatory, 520 Edgemont Rd, Charlottesville, VA 22903, USA}

\author[0000-0003-2141-5689]{Peter Schilke}
\affiliation{I. Physikalisches Institut der Universit{\"a}t zu K{\"o}ln, Z{\"u}lpicher Str. 77, 50937, K{\"o}ln, Germany}

\author[0000-0002-1069-2931]{Emmanuel Garcia-Berrios}
\affiliation{Astrochemistry Laboratory, NASA Goddard Space Flight Center, Greenbelt, MD 20771, USA}
\affiliation{Institute for Astrophysics and Computational Sciences, The Catholic University of America, Washington, DC 20064, USA}
\affiliation{Department of Astronomy, University of Illinois, 1002 W. Green St., Urbana, IL 61801, USA}

\begin{abstract}
We report the first detection of deuterated water (HDO) toward an extragalactic hot core. The HDO  2$_\mathrm{11}$--2$_\mathrm{12}$ line has been detected toward hot cores N\,105--2\,A and 2\,B in the N\,105 star-forming region in the low-metallicity Large Magellanic Cloud (LMC) dwarf galaxy with the Atacama Large Millimeter/submillimeter Array (ALMA). We have compared the HDO line luminosity ($L_{\rm HDO}$) measured toward the LMC hot cores to those observed toward a sample of seventeen Galactic hot cores covering three orders of magnitude in $L_{\rm HDO}$, four orders of magnitude in bolometric luminosity ($L_{\rm bol}$), and a wide range of Galactocentric distances (thus metallicities).  The observed values of $L_{\rm HDO}$ for the LMC hot cores fit very well into the $L_{\rm HDO}$ trends with $L_{\rm bol}$ and metallicity observed toward the Galactic hot cores. We have found that $L_{\rm HDO}$ seems to be largely dependent on the source luminosity, but metallicity also plays a role.  We provide a rough estimate of the H$_2$O column density and abundance ranges toward the LMC hot cores by assuming that HDO/H$_2$O toward the LMC hot cores is the same as that observed in the Milky Way; the estimated ranges are systematically lower than Galactic values. The spatial distribution and velocity structure of the HDO emission in N\,105--2\,A is consistent with HDO being the product of the low-temperature dust grain chemistry. Our results are in agreement with the astrochemical model predictions that HDO is abundant regardless of the extragalactic environment and should be detectable with ALMA in external galaxies. 
\end{abstract}

\section{Introduction}
\label{s:intro}

\begin{figure*}[ht!]
\centering
\includegraphics[width=0.48\textwidth]{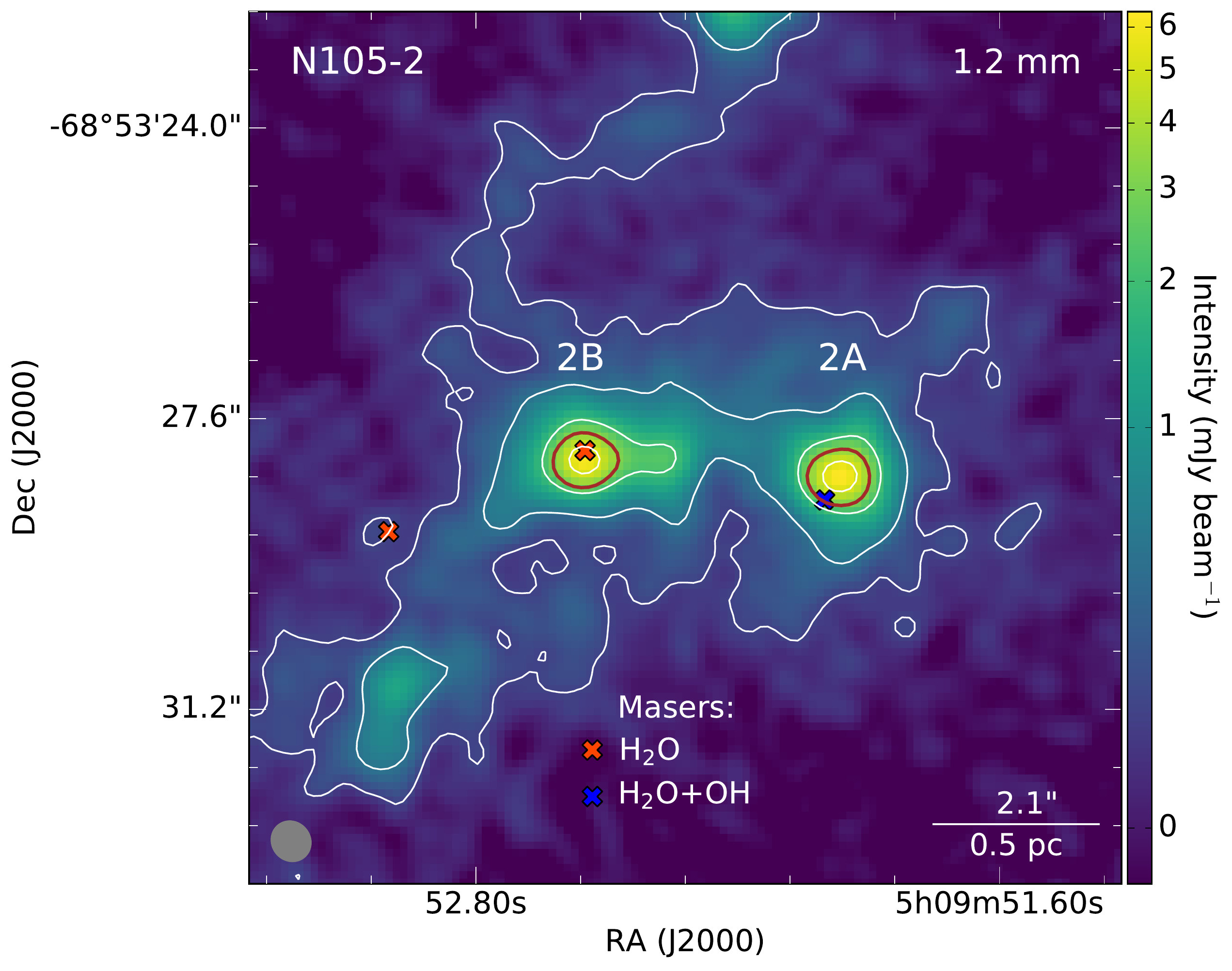}
\includegraphics[width=0.49\textwidth]{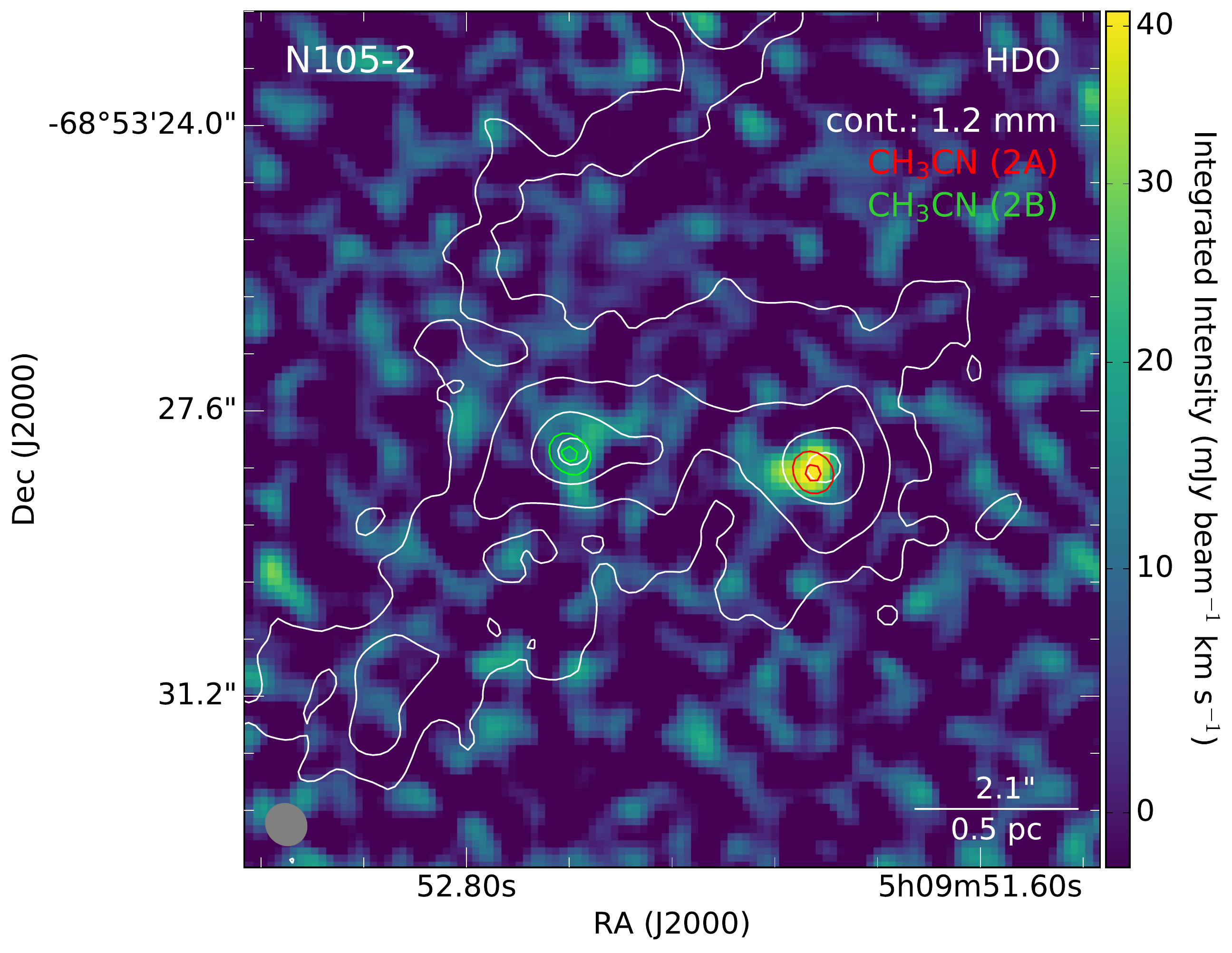}
\caption{{\it Left}: The 1.2 mm continuum image of N\,105--2 with spectral extraction regions for 2\,A and 2\,B indicated in brown; they are contours corresponding to 50\% of the continuum peak of the corresponding source (6.4 and 6.2 mJy beam$^{-1}$ for 2\,A and 2\,B, respectively). {\it Right}:  The HDO 2$_{11}$--2$_{12}$ integrated intensity image (moment 0 map) of N\,105--2 for a velocity range from 240.4 km~s$^{-1}$ to 250.7 km~s$^{-1}$.  Red/green contours correspond to (50, 90)\% of the CH$_3$CN integrated intensity peak of 0.38/0.19 Jy beam$^{-1}$ km~s$^{-1}$ for 2\,A/2\,B. White contours in both panels represent the 1.2 mm continuum emission with contour levels of (3, 10, 40, 100) $\times$ the image rms noise level ($\sigma$) of $5.1\times10^{-5}$ Jy beam$^{-1}$.  The positions of H$_2$O and OH masers are indicated in the left panel. The size of the synthesized beam is shown in the lower left corner of each image. \label{f:cont}}
\end{figure*}

Water (H$_2$O) is a key molecule tracing the chemical and physical processes associated with the formation of stars and planets. Water shows large abundance variations in star-forming regions because it can be produced in both gas-phase and on the surfaces of interstellar dust grains (e.g., \citealt{vandishoeck2021}). In cold molecular gas, most water is in the form of ice, with only trace amounts in the gas. In outflow shocks where $T > 300$ K, water is predominantly in the gas phase where it forms directly (e.g., \citealt{suutarinen2014}; \citealt{kristensen2017}; \citealt{karska2018}). Deuterated water (HDO), on the other hand, forms mostly on the dust grains in the cold clouds before core collapse (e.g., \citealt{jacq1990}; \citealt{furuya2016}). Particularly, the amount of HDO formed is set by a combination of the temperature and life-time of the cold phase, where higher temperatures and shorter life-times lead to lower deuterium fractionation, and vice versa (e.g., \citealt{jensen2021}). Once formed on the grains, HDO typically sublimates into the gas phase near protostars, where the dust temperature exceeds 100 K, in so-called hot cores (high-mass stars) or hot corinos (low- and intermediate-mass stars; e.g., \citealt{herbst2009}). The amount of HDO present thus contains a fossil record of the conditions in the cold gas, and a key question naturally arises:  how will different physical conditions in external galaxies affect these processes? 

The first, and until now the only, extragalactic detection of HDO was reported by \citet{muller2020}.  Using the Atacama Large Millimeter/submillimeter Array (ALMA), \citet{muller2020} detected the HDO $J_{\rm K_a, K_c}$ = $1_{01}$--$0_{00}$ absorption line at 464.9245 GHz in a spiral galaxy at a redshift ($z$) of 0.89 on the line of sight toward the quasar PKS 1830$-$211. Here, we report the first detection of HDO toward extragalactic hot molecular cores.  Hot cores are compact ($\lesssim$0.1 pc), warm ($\gtrsim$100 K), and dense ($\gtrsim$10$^{6-7}$ cm$^{-3}$) regions surrounding high-mass protostars very early in their evolution. A typical Galactic hot core is chemically rich, containing the products of the interstellar grain-surface chemistry (including complex organics and water) released from the dust grain ice mantles to the gas phase via thermal evaporation and/or sputtering in shock waves (e.g., \citealt{garay1999}; \citealt{kurtz2000}; \citealt{cesaroni2005}; \citealt{palau2011}). Hot cores may also display products of post-desorption gas chemistry (e.g., \citealt{herbst2009}; \citealt{oberg2016}; \citealt{jorgensen2020}). 

We detected the HDO  2$_{11}$--2$_{12}$ line at 241.5616 GHz with ALMA toward hot cores N\,105--2\,A and N\,105--2\,B in the star-forming region N\,105 in the Large Magellanic Cloud (LMC; briefly reported in \citealt{sewilo2022}). These are two out of only a handful of known bona fide extragalactic hot cores, all located in the LMC (\citealt{shimonishi2016b,shimonishi2020}; \citealt{sewilo2018,sewilo2019,sewilo2022})

The LMC, an irregular dwarf galaxy, is the most massive and one of the nearest ($50.0\pm1.1$ kpc;  \citealt{pietrzynski2013}) satellites of the Milky Way.  The low metallicity of the LMC ($Z$$\sim$0.3--0.5 $Z_{\odot}$; \citealt{russell1992}; \citealt{westerlund1997}; \citealt{rolleston2002}), similar to galaxies at the peak of star formation in the Universe ($z$$\sim$1.5; e.g., \citealt{pei1999}, \citealt{mehlert2002}; \citealt{madau2014}), provides a unique opportunity to study star formation (including the H$_2$O and HDO chemistry) in an environment which is significantly different than in today's Galaxy. 

There are several factors that can directly impact the formation and destruction of H$_2$O and HDO molecules in a low-metallicity environment.  The abundance of atomic O in the LMC is over a factor of two lower when compared with the Galaxy (i.e., fewer O atoms are available for water chemistry; e.g., \citealt{russell1992}).   The dust-to-gas ratio in the LMC is lower (e.g., \citealt{dufour1975,dufour1984}; \citealt{koornneef1984}; \citealt{duval2014}), resulting in fewer dust grains for surface chemistry and less shielding than in the Galaxy.  The deficiency of dust combined with the harsher UV radiation field in the LMC (e.g., \citealt{browning2003}; \citealt{welty2006}) leads to warmer dust temperatures (e.g., \citealt{vanloon2010smc}) and consequently, less efficient grain-surface reactions (e.g., \citealt{shimonishi2016a}; \citealt{acharyya2015}).  The cosmic-ray density in the LMC is about 25\% of that measured in the solar neighborhood  (e.g., \citealt{abdo2010}; \citealt{knodlseder2013}), resulting in less effective cosmic-ray-induced UV radiation. 

\begin{figure*}[ht!]
\centering
\includegraphics[width=\textwidth]{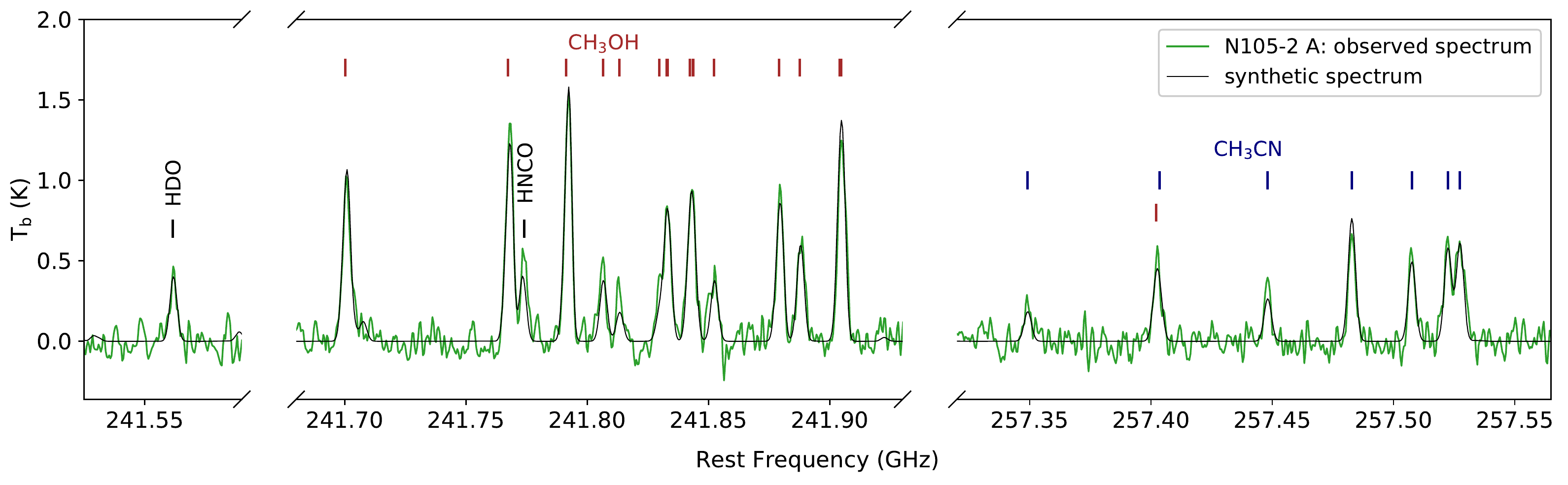}
\includegraphics[width=\textwidth]{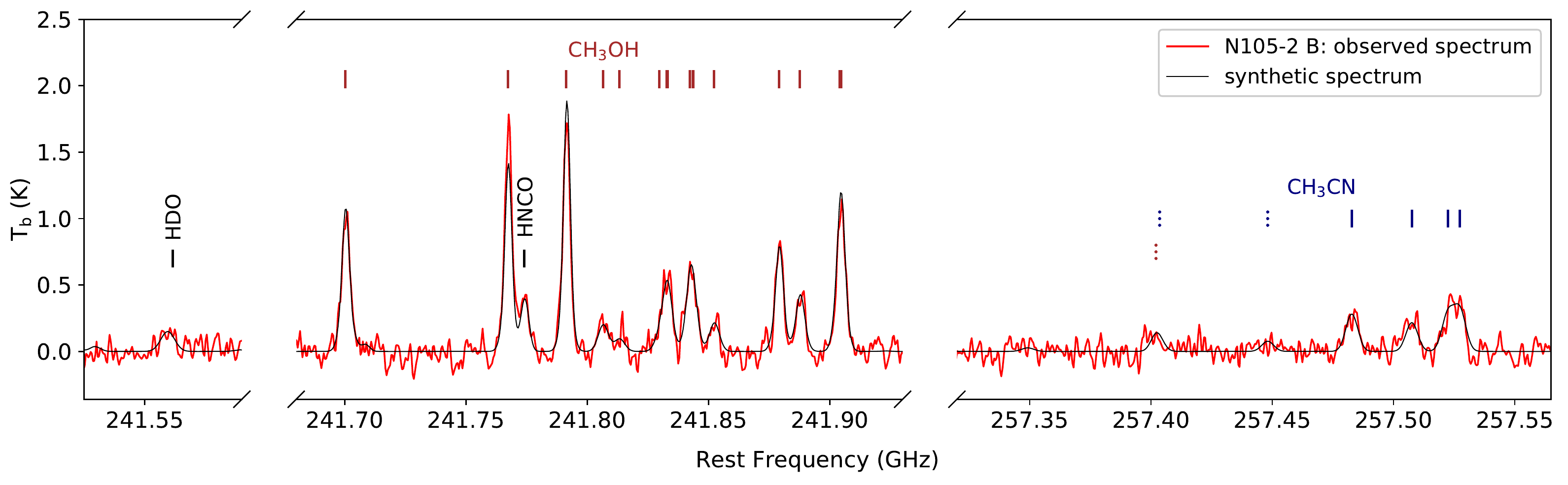}
\caption{Selected frequency ranges of the ALMA Band 6 spectra of hot cores N\,105--2\,A ({\it top}) and 2\,B ({\it bottom}), covering the HDO 2$_\mathrm{11}$--2$_\mathrm{12}$ line, as well as the CH$_3$OH $J$ = 5--4 Q-branch and the CH$_3$CN 14$_K$--13$_K$ ladder for reference.  The tentatively detected transitions are indicated with dotted lines.  The synthetic spectra from Sewi{\l}o et al. (2022) are shown in black. \label{f:spec}}
\end{figure*}

Extragalactic deuterated molecules were first detected in star-forming regions of the LMC by \citet{chin1996} in single-dish observations.  Deuterated formyl cation (DCO$^{+}$) was detected toward three (N\,113, N\,44\,BC, N\,159\,HW) and deuterated hydrogen cyanide (DCN) toward one star-forming region (N\,113;  see also \citealt{wang2009}). In an independent study, \citet{heikkila1997} reported a detection of DCO$^{+}$ and a tentative detection of DCN toward N\,159.  

In the more recent interferometric studies, deuterated molecules have been detected toward the LMC hot cores and hot core candidates. DCN was detected in two hot cores in N\,113 (N\,113\,A1 and N\,113\,B3; \citealt{sewilo2018}),  deuterated hydrogen sulfide (HDS) toward a hot core candidate N\,105--2\,C, and deuterated formaldehyde (HDCO) and HDO toward hot cores N\,105--2\,A and N\,105--2\,B, (\citealt{sewilo2022}).  In this paper, we provide a detailed discussion on the detection of HDO toward N\,105--2\,A and 2\,B: the first detection of HDO toward an extragalactic hot core.

\section{The Data}
\label{s:data}

Field N\,105--2 in the star-forming region LHA\,120--N\,105 (hereafter N\,105; \citealt{henize1956}) hosting hot cores 2\,A and 2\,B was observed with ALMA 12 m Array in Band 6 as part of the Cycle 7 project 2019.1.01720.S (PI M. Sewi{\l}o; \citealt{sewilo2022}). The observations were executed twice on October 21, 2019 with 43 antennas and baselines from 15 m to 783 m.  The (bandpass, flux, phase) calibrators were (J0519$-$4546, J0519$-$4546, J0440$-$6952) and (J0538$-$4405, J0538$-$4405, J0511$-$6806) for the first and second run, respectively. N\,105--2 was observed again on October 23, 2019 with 43 antennas, baselines from 15 m to 782 m, and the same calibrators.  The total on-source integration was $\sim$13.1 minutes.  The spectral setup included four 1875 MHz spectral windows with 3840 channels centered on frequencies of 242.4 GHz, 244.8 GHz, 257.85 GHz, and 259.7 GHz; the spectral resolution is 1.21--1.13 km~s$^{-1}$.  

The data were calibrated and imaged with version 5.6.1-8 of the ALMA pipeline in CASA (Common Astronomy Software Applications; \citealt{mcmullin2007}).  Continuum was subtracted in the uv domain from the line spectral windows. The CASA task \texttt{tclean} was used for imaging using the Hogbom deconvolver, standard gridder, Briggs weighting with a robust parameter of 0.5, and auto-multithresh masking.   The spectral cubes have a cell size of  $0\rlap.{''}092 \times 0\rlap.{''}092 \times 0.56$ km s$^{-1}$ and they have been corrected for primary beam attenuation. 

Here, we present the results based on the 242.4 GHz spectral window: a detection of the HDO 2$_{11}$--2$_{12}$ transition at 241,561.550 MHz with the upper energy level $E_{\rm U}$ of 95.2 K toward two continuum sources (A and B) in the N\,105--2 field (see Fig.\ref{f:cont}). Sensitivity of 1.97 mJy  per $0\rlap.{''}54\times0\rlap.{''}50$ beam (0.15 K) was achieved in the 242.4 GHz cube.  Sensitivity of 0.05 mJy per $0\rlap.{''}51\times0\rlap.{''}$47 beam  (4.4 mK) was achieved in the continuum.  

\section{Results}
\label{s:results} 

Figure \ref{f:spec} shows selected frequency ranges of the ALMA Band 6 spectra of hot cores N\,105--2\,A and B, covering the HDO 241.6 GHz line, as well as the methanol (CH$_3$OH) $J$ = 5--4 Q-branch at $\sim$241.8 GHz and the methyl cyanide (CH$_3$CN) 14$_K$--13$_K$ ladder for reference.  The molecular line identification and spectral modeling for all spectral windows were performed for all the continuum sources in N\,105--2 in \citet{sewilo2022}.  

The spectral line modeling was performed using a least-squares approach under the assumption of local thermodynamic equilibrium (LTE) and accounting for line opacity effects.  The best-fitting column density, rotational temperature, Doppler shift, and spectral line width ([$N^i, T_{\rm rot}^i, v_i, dv^i$]) for the complete set of species were determined simultaneously.  A custom Python routine was used to generate spectral line models with spectroscopic  parameters taken from the Cologne Database for Molecular Spectroscopy (CDMS\footnote{http://www.astro.uni-koeln.de/cdms},  \citealt{muller2001}) for all molecular species except HDO (not included in CDMS) for which the data were taken from the Jet Propulsion Laboratory (JPL) Molecular Spectroscopy Database\footnote{http://spec.jpl.nasa.gov/} (\citealt{pickett1998}).  For molecular species with single line detections (including HDO), the rotational temperature of CH$_3$CN, $T_{\rm CH_3CN}$,  was adopted for the fitting. For N\,105--2\,A and 2\,B, the HDO and CH$_3$CN integrated intensity peaks coincide, supporting this assumption (see Figs.~\ref{f:cont} and \ref{f:water}). 

The \citet{sewilo2022}'s LTE spectral fitting results for HDO ([$N, v, dv$] = [$N_{\rm HDO}, v_{\rm LSR}, \Delta v_{\rm FWHM}$]) and the adopted $T_{\rm CH_3CN}$ are provided in Table~\ref{t:phys} for N\,105--2\,A and 2\,B. The synthetic spectra are overlaid on the observed spectra of 2\,A and 2\,B in Fig.~\ref{f:spec}. Table~\ref{t:phys} also lists the H$_2$ column densities ($N_{\rm H_2}$),  H$_2$ number densities ($n_{\rm H_2}$), and HDO abundances with respect to H$_2$ ($N_{\rm HDO}$/$N_{\rm H_2}$).  $N_{\rm H_2}$ was calculated by \citet{sewilo2022} based on the 1.2 mm continuum flux density and adopting $T_{\rm CH_3CN}$ under the assumption that the dust and gas are well-coupled. The assumption of thermal equilibrium between the dust and gas holds for high-density regions such as N\,105--2\,A and 2\,B ($n_{\rm H_2} \gtrsim 10^5$ cm$^{-3}$; e.g., \citealt{goldsmith1978}; \citealt{kaufman1998}). 

We have measured the HDO 2$_\mathrm{11}$--2$_\mathrm{12}$ line flux (the integrated line intensity; $F_{\rm HDO}$) of $1.8\pm0.3$ K km~s$^{-1}$ for N\,105--2\,A and $1.3\pm0.6$ K km~s$^{-1}$ for 2\,B. We have calculated the HDO 2$_\mathrm{11}$--2$_\mathrm{12}$ line luminosity ($L_{\rm HDO}$) from $F_{\rm HDO}$ using the standard relation (e.g., \citealt{wu2005}) as outlined in Appendix~\ref{s:appA}.  $L_{\rm HDO}$ is $(3.0\pm0.5)\times10^{-2}$ $L_{\odot}$ for 2\,A and $(2.2\pm1.0)\times10^{-2}$ $L_{\odot}$ for 2\,B.  The results are listed in Table~\ref{t:thedata}.

 \begin{deluxetable}{lcc}
\centering
\tablecaption{A Summary of the Physical Properties of the LMC Hot Cores with the HDO Detection N\,105--2\,A and 2\,B \citep{sewilo2022}  \label{t:phys}} 
\tablewidth{0pt}
\tablehead{
\colhead{Parameter} &
\colhead{N\,105--2\,A} &
\colhead{N\,105--2\,B} 
}
\startdata
$T_{\rm CH_3CN}$ (K) & $152^{+10}_{-11}$ & $88^{+10}_{-9}$ \\
$N_{\rm HDO}$  (cm$^{-2}$)  & $(4.9^{+0.5}_{-0.4})\times10^{14}$ & $(2.6\pm0.5)\times10^{14}$\\
$v_{\rm LSR}$ (km s$^{-1}$) & $242.7\pm0.2$ & $245.8^{+0.8}_{-0.7}$\\
$\Delta v_{\rm FWHM}$ (km s$^{-1}$)  & $4.2\pm0.4$ & $8.2^{+2.3}_{-1.8}$\\
$N_{\rm H_2}$  (cm$^{-2}$)  & $(1.8\pm0.2)\times10^{23}$ & $(3.1\pm0.5)\times10^{23}$\\
$n_{\rm H_2}$ (cm$^{-3}$)  & $\sim$$4.6\times10^5$ & $\sim$$7.8\times10^5$\\
$N_{\rm HDO}$/$N_{\rm H_2}$ & $(2.7\pm0.4)\times10^{-9}$ & $(8.2^{+2.1}_{-2.0})\times10^{-10}$ \\
\enddata
\end{deluxetable}

\subsection{The Galactic Sample of Hot Cores with the HDO Detection}
\label{s:galsample}

 HDO observations are available in the literature for seventeen Galactic hot cores. The HDO 2$_\mathrm{11}$--2$_\mathrm{12}$ line fluxes (same transition we detected in the LMC) are available for W3(H$_2$O), AFGL\,2591, G34.26$+$0.15, W51\,e1/e2, W51\,d, NGC\,7538\,IRS1,  Sgr\,B2(N) and Sgr\,B2(M) near the Galactic Center, and the extreme outer Galaxy source WB\,89--789 SMM1.

Observations of W3(H$_2$O) were performed with the James Clerk Maxwell Telescope (JCMT) with a 19$\rlap.{''}$7 beam (half-power beam width, HPBW), tracing 0.2 pc linear scales  \citep{helmich1996}.  The HDO data for  AFGL\,2591 \citep{vandertak2006}, G34.26$+$0.15 \citep{coutens2014},  W51\,e1/e2, W51\,d, and NGC\,7538\,IRS1  \citep{jacq1990}, were obtained with the IRAM 30m telescope with a 12$''$ beam, tracing 0.15--0.32 pc scales for the distance range covered by these sources.  The Sgr\,B2(N) and Sgr\,B2(M) observations were performed with the SEST telescope with a 22$''$ beam, tracing 0.89 pc scales \citep{nummelin2000}. WB\,89--789 SMM1 was observed with ALMA by \citet{shimonishi2021} with a $\sim$0$\rlap.{''}$5 beam, corresponding to $\sim$0.026 pc.

\begin{figure*}[ht!]
\centering
\includegraphics[width=\textwidth]{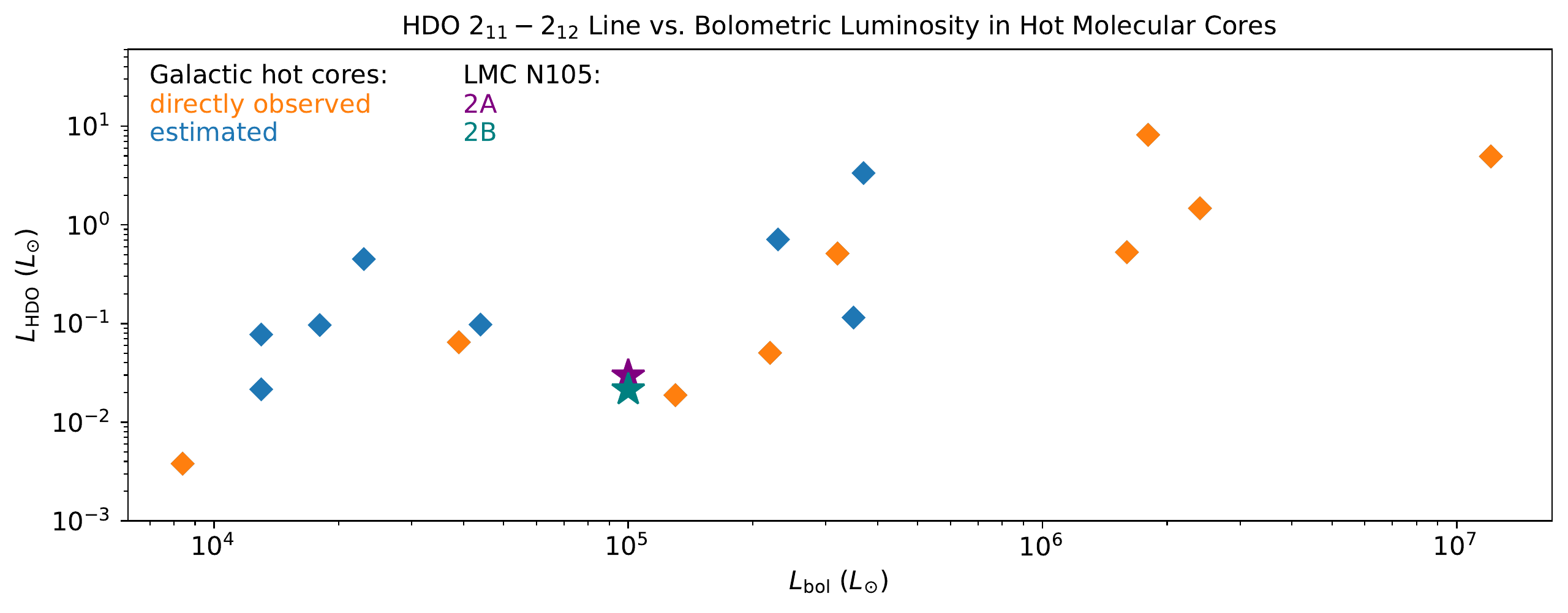} 
\caption{The HDO 241.6 GHz line luminosities ($L_{\rm HDO}$) measured toward hot cores N\,105--2\,A and 2\,B in the LMC and those observed toward a sample of hot cores in the Milky Way as a function of the bolometric luminosity ($L_{\rm bol}$). The Galactic hot cores shown in the plot are (in order of increasing $L_{\rm bol}$): WB\,89--789 SMM1 (an extreme outer Galaxy source), NGC 7538 S and IRAS 18089$-$1732 ($L_{\rm HDO}$=0.02 $L_{\odot}$ and 0.08 $L_{\odot}$, respectively), G9.62$+$0.19, W43 MM1, W3(H$_2$O),  W33A, NGC 7538 IRS1, AFGL 2591, G31.41$+$0.31, G34.26$+$0.15,  G29.96$-$0.02, G10.47$+$0.03A, W51 e1/e2, Sgr B2(N),  W51d, Sgr B2(M).  The values of $L_{\rm HDO}$ indicated with orange diamonds are based on the observations of the HDO $2_{11}-2_{12}$ transition, while those indicated with blue diamonds were estimated using the observations of other HDO transitions as described in Section~\ref{s:results}.   \label{f:lmcgal}}
\end{figure*}

Two HDO transitions were detected toward W43\,MM1, NGC\,7538\,S, IRAS\,18089$-$1732 \citep{marseille2010}, and W33A \citep{vandertak2006} with the IRAM 30m telescope:  1$_\mathrm{10}$--1$_\mathrm{11}$ (80.5783 GHz, $E_{\rm up}$ = 46.8 K; 30$''$ beam, 0.34--0.80 pc scales) and 3$_\mathrm{12}$--2$_\mathrm{21}$ (225.8967 GHz, $E_{\rm up}$ = 167.6 K; 11$''$ beam, 0.12--0.31 pc scales).  Assuming that these two transitions are optically thin and in LTE (see e.g., \citealt{persson2014}), we have used a rotational diagram (\citealt{Goldsmith1999}) to estimate the HDO 2$_\mathrm{11}$--2$_\mathrm{12}$ line flux toward W43\,MM1, NGC\,7538\,S, IRAS\,18089$-$1732, and W33A (see Appendix~\ref{s:appA}).

Data for a single HDO transition, 3$_\mathrm{12}$--2$_\mathrm{21}$, are available for G9.62$+$0.19, G10.47$+$0.03A, G29.96$-$0.02, G31.41$+$0.31 \citep{gensheimer1996}; the IRAM 30m telescope observations of these sources trace 0.2--0.6 pc scales. To estimate the HDO 2$_\mathrm{11}$--2$_\mathrm{12}$ line flux, we extrapolated the 3$_\mathrm{12}$--2$_\mathrm{21}$ line flux assuming an excitation temperature derived in literature for these sources using CH$_3$CN: 70 K for G9.62$+$0.19 \citep{hofner1996}, 164 K for G10.47$+$0.03A \citep{olmi1996}, 160 K for G29.96$-$0.02 \citep{beltran2011}, and 158 K for G31.41$+$0.31 \citep{beltran2005}. The calculated values of the HDO 2$_\mathrm{11}$--2$_\mathrm{12}$ line flux for G9.62$+$0.19, G10.47$+$0.03A, G29.96$-$0.02, and G31.41$+$0.31 are the most uncertain of all  Galactic sources in our sample.  However, in Appendix~\ref{s:appA}, we show that the results for G9.62$+$0.19, G10.47$+$0.03A, G29.96$-$0.02, and G31.41$+$0.31 do not change significantly when different values of temperature are adopted (60--200 K).

We have derived the HDO 2$_\mathrm{11}$--2$_\mathrm{12}$ line luminosities from line fluxes for Galactic hot cores using the same formula as for N\,105--2\,A and 2\,B.  The value of $L_{\rm HDO}$  spans three orders of magnitude, ranging from $3.8\times10^{-3}$ $L_{\odot}$ for WB\,89--789 SMM1 to 8.2 $L_{\odot}$ for Sgr\,B2(N). Both the HDO 2$_\mathrm{11}$--2$_\mathrm{12}$ line fluxes and luminosities for Galactic hot cores analyzed in this paper are provided in Table~\ref{t:thedata} in Appendix~\ref{s:appA}. 

Our ALMA observations of a star-forming region in the LMC at $\sim$50 kpc with a resolution of $\sim$0$\rlap.{''}$5 probe physical scales of 0.12--0.13 pc, similar to those traced by the observations of Galactic sources with single-dish telescopes such as the IRAM 30m at 241.6 GHz, at a distance of $\sim$2 kpc.

\begin{figure*}[ht!]
\centering
\includegraphics[width=0.85\textwidth]{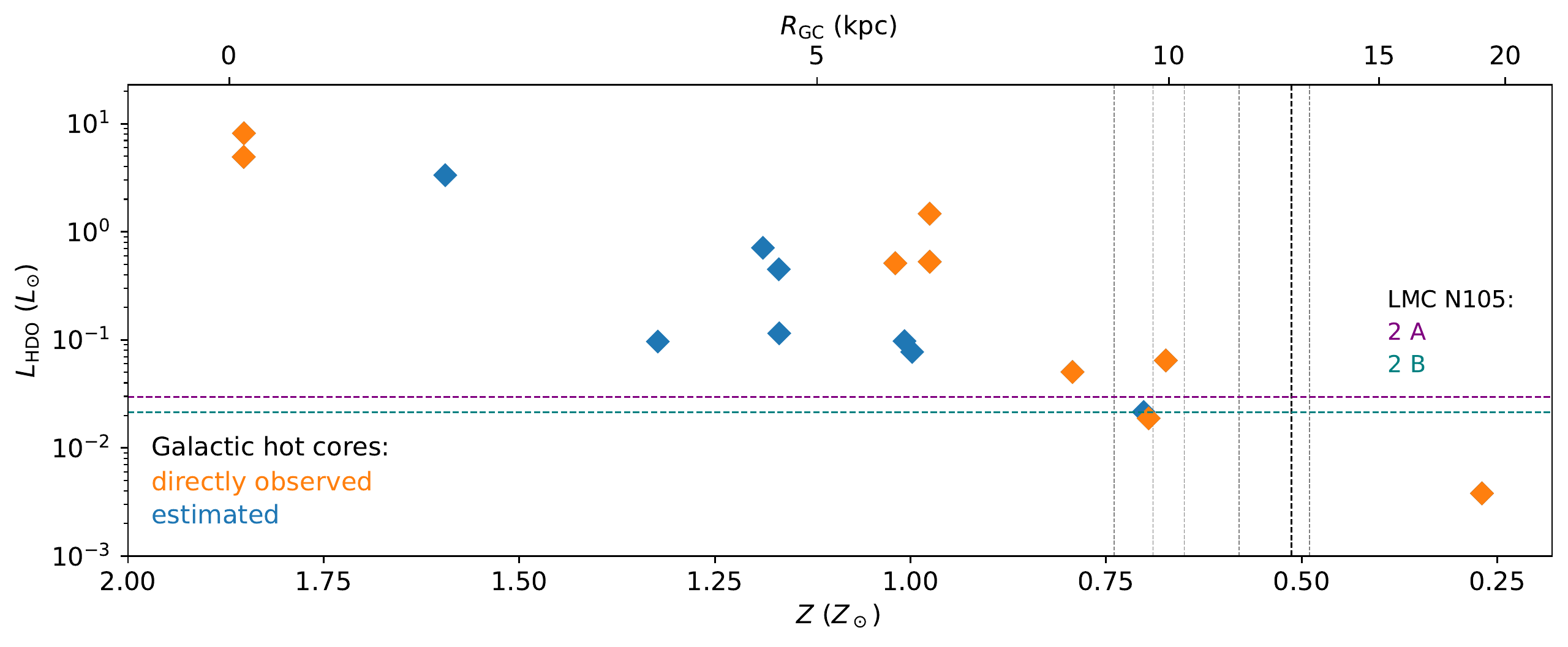}
\includegraphics[width=0.85\textwidth]{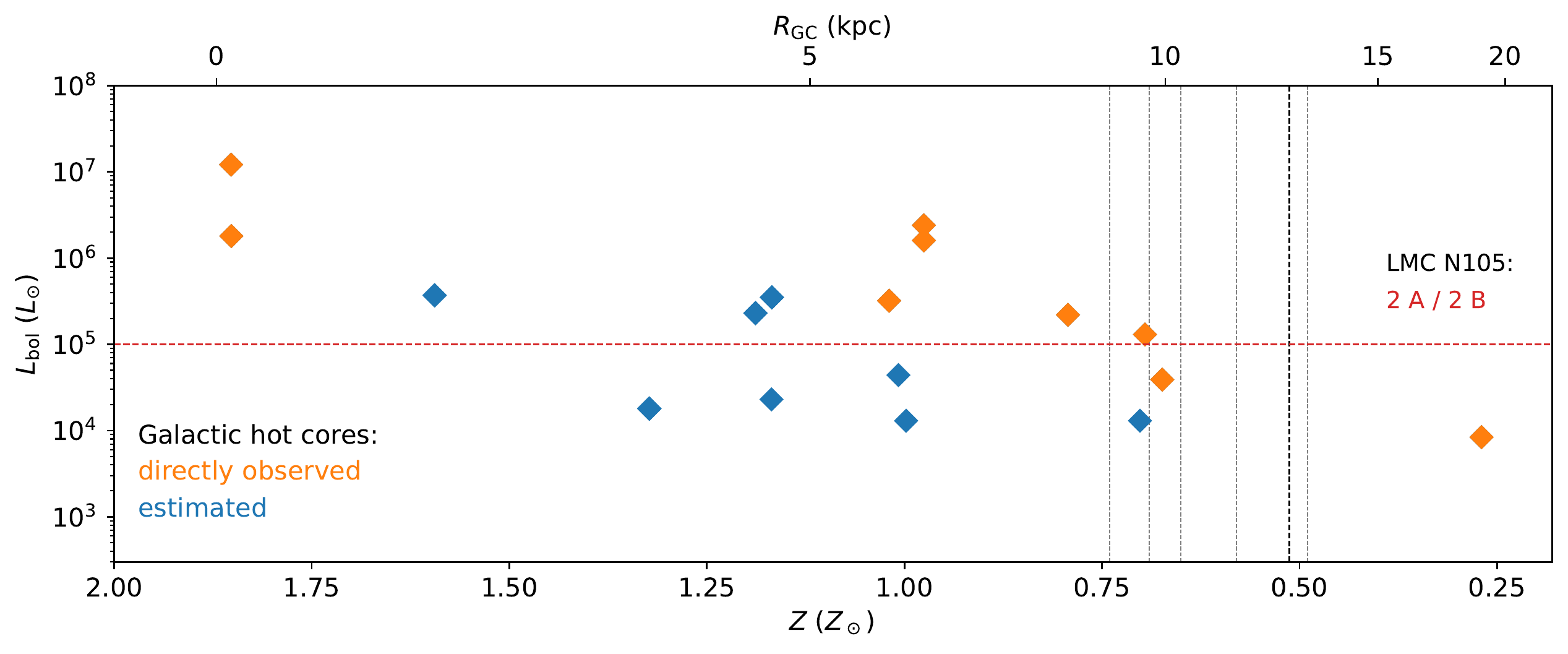}
\includegraphics[width=0.85\textwidth]{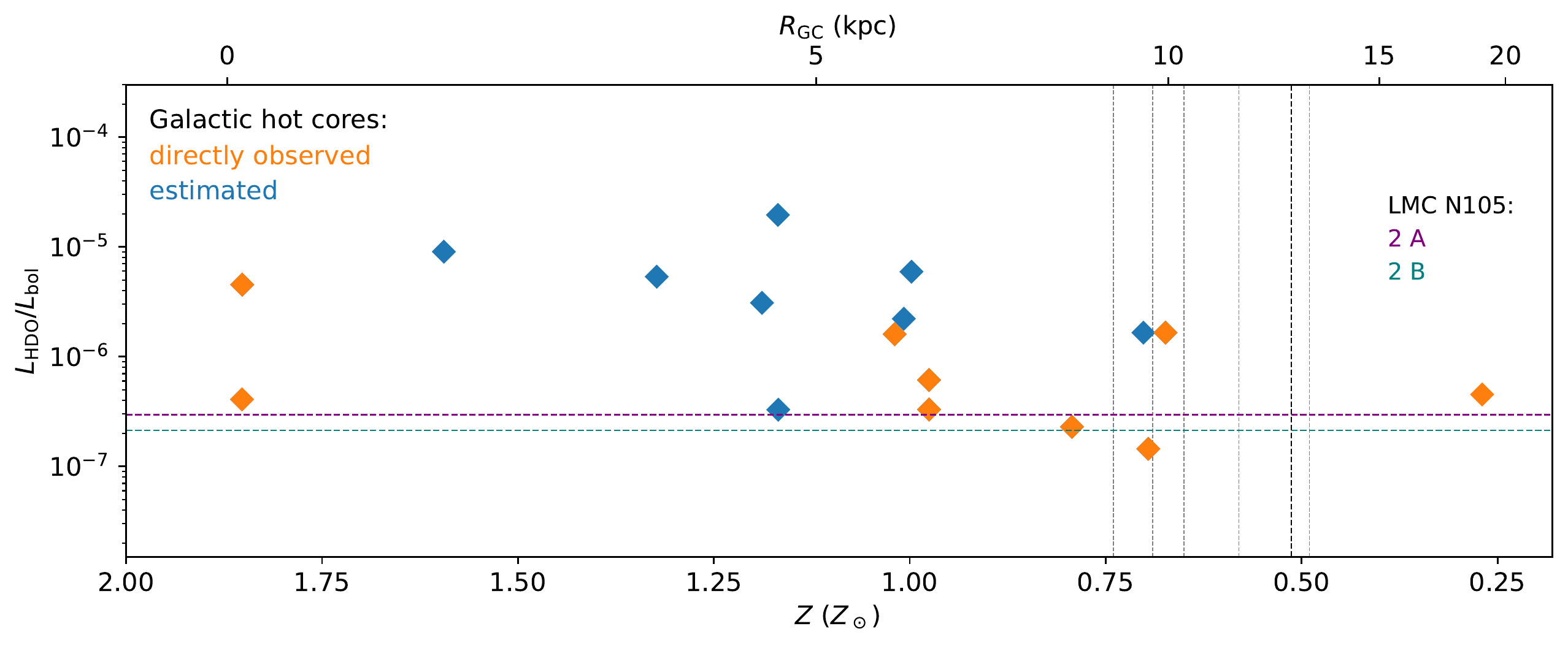}
\caption{The HDO 241.6 GHz line luminosities ($L_{\rm HDO}$; upper panel), bolometric luminosities ($L_{\rm bol}$; middle panel),  and the $L_{\rm HDO}$/$L_{\rm bol}$ ratio (lower panel) of the Galactic hot cores as a function of the Galactocentric distance ($R_{\rm GC}$) and metallicity ($Z$).  The metallicity at a given $R_{\rm GC}$ was calculated using the \citet{balser2011}'s  O/H radial gradient $12+{\rm log(O/H)}=-0.0446\,R_{\rm GC}+8.962$ and adopting [$12+{\rm log(O/H)]_{\odot}}$ of 8.69 (\citealt{asplund2009}): ${\rm log}(Z/Z_\odot)={\rm log(O/H)}-{\rm log(O/H)}_{\odot}$ The purple/teal solid lines in the upper panel indicate the measured $L_{\rm HDO}$ for the LMC hot cores N\,105--2\,A/2\,B, while the red line in the lower panel corresponds to their roughly equal $L_{\rm bol}$. The vertical black line indicates $R_{\rm GC}$ where $[12+{\rm log(O/H)}]=[12+{\rm log(O/H)}]_{\rm LMC}=8.4$ (e.g., \citealt{russell1992}). The vertical gray lines indicate $R_{\rm GC}$ where $[12+{\rm log(O/H)}]=8.4$ based on the O/H gradients found in other H\,{\sc ii} region studies, from left to right:  \citet{rudolph2006}, far-IR data; \citet{esteban2018}; \citet{arellano2020}; \citet{fernandez2017}; and \citet{rudolph2006}, optical data. The orange and blue symbols are the same as in Fig.~\ref{f:lmcgal}. \label{f:lmcgal2} 
}
\end{figure*}

\section{Discussion}
\label{s:discussion} 

\subsection{HDO 2$_\mathrm{11}$--2$_\mathrm{12}$ Line Luminosity: LMC vs. Galactic Hot Cores}
\label{s:compLgal} 

Figure~\ref{f:lmcgal} shows the HDO 2$_\mathrm{11}$--2$_\mathrm{12}$ line luminosities ($L_{\rm HDO}$) measured toward Galactic and LMC hot cores as a function of bolometric luminosity ($L_{\rm bol}$).  $L_{\rm bol}$ of Galactic hot cores ranges from $8.4\times10^{3}$ $L_{\odot}$  for WB\,89--789 SMM1 to $1.2\times10^{7}$ $L_{\odot}$ for Sgr\,B2(M). The values of $L_{\rm bol}$ were adopted from \citet{shimonishi2021} for WB\,89--789 SMM1, \citet{wright2012} for NGC\,7538\,S, \citet{hofner1996} for G9.62$+$0.19, \citet{ahmadi2018} for W3(H$_2$O), \citet{hernandez2014} and \citet{vandertak2013} for W51\,e1/e2, \citet{rolffs2011} for W51\,d, \citet{schmiedeke2016} for Sgr\,B2(N) and Sgr\,B2(M),  and \citet{vandertak2013} for the remaining sources. 

There are uncertainties in $L_{\rm bol}$  related to a relatively low resolution of the single-dish observations. For example, the bolometric luminosities of G29.96$-$0.02 and G34.26$+$0.15 likely include contributions from both hot cores and nearby ultracompact (UC) H\,{\sc ii} regions. The HDO emission toward both regions was detected with the IRAM 30m telescope and thus all of these components were within the half-power beam width.  We do, however, expect most of the HDO emission to come from hot cores rather than more evolved UC H\,{\sc ii} regions. 

Insufficient multi-wavelength high-resolution data are available to determine individual $L_{\rm bol}$ for the LMC hot cores N\,105--2\,A and 2\,B.  To make an estimate of their $L_{\rm bol}$, we determined a combined $L_{\rm bol}$ based on the data from 3.6 $\mu$m to 1.2 mm and inferred a contribution from each source as described in Appendix~\ref{s:lbol}. We estimate that both N\,105--2\,A and 2\,B have $L_{\rm bol}$ of $\sim$10$^{5}$ $L_{\odot}$.  Since the sample of Galactic hot cores used for the analysis covers a wide range of the Galactocentric distances (thus metallicities; see below), we did not apply a correction to $L_{\rm HDO}$ measured toward the LMC hot cores to account for a difference in the metallicity between the LMC and the solar neighborhood.

The trend of increasing $L_{\rm HDO}$ with increasing $L_{\rm bol}$ for Galactic hot cores is very suggestive in Fig.~\ref{f:lmcgal}, especially when only the direct measurements of the HDO transition detected in the LMC are taken into account.  The observed values of $L_{\rm HDO}$ for the LMC hot cores N\,105--2\,A and 2\,B fit into this trend very well.  The higher abundance of HDO for more luminous young stellar objects with hot cores is expected since the higher temperatures result in more HDO to be released from the icy grain mantles in hot core regions. $L_{\rm HDO}$ is also expected to scale with the total HDO column density which can be affected by low metallicity, a lower atomic O abundance in particular. 

We can test the dependence of $L_{\rm HDO}$ on metallicity by investigating how $L_{\rm HDO}$ changes as a function of the Galactocentric distance ($R_{\rm GC}$).  The observations of a variety of objects including H\,{\sc ii} regions and Cepheid variable stars revealed radial elemental abundance gradients in the Milky Way disk (e.g., \citealt{churchwell1975}; \citealt{maciel2019} and references therein).  Traced by O/H and Fe/H, metallicity decreases with increasing $R_{\rm GC}$.  

The O/H gradients based on Cepheids have slopes between $-$0.05 dex/kpc and $-$0.06 dex/kpc; similar slopes within the uncertainties have been obtained for the Fe/H gradients (e.g., \citealt{maciel2019} for over 300 Cepheids and $R_{\rm GC}$$\sim$3--18 kpc). The O/H gradients from much smaller samples of H\,{\sc ii} regions are also similar to those measured from Cepheids within the uncertainties, ranging from $-$0.04 dex/kpc to $-$0.06 dex/kpc (e.g., \citealt{fernandez2017} and references therein; \citealt{esteban2018}). 

We calculated $R_{\rm GC}$ for Galactic hot cores shown in Fig.~\ref{f:lmcgal} based on their Galactic coordinates and distances (kinematic or parallax), and assuming the distance to the Galactic Center of 8.34 kpc \citep{reid2014}. Located near the Galactic Center, Sgr\,B2(N) and Sgr\,B2(M) hot cores represent a high-metallicity environment ($Z_{\odot}$ $<$ $Z_{\rm GC}$ $\lesssim$ 2 $Z_{\odot}$; \citealt{schultheis2019} and references therein) and have the highest $L_{\rm HDO}$, while the extreme outer Galaxy source WB\,89--789 SMM1 with the lowest $L_{\rm HDO}$ is in the low-metallicity environment ($\sim$0.25  $Z_{\odot}$).  In general, with increasing $R_{\rm GC}$ and thus decreasing O/H ratio (metallicity), $L_{\rm HDO}$ decreases  (see the top panel in Fig.~\ref{f:lmcgal2}). 

Four Galactic hot cores with $L_{\rm HDO}$ most similar to that measured toward N\,105--2\,A and 2\,B (NGC\,7538\,S, NGC\,7538\,IRS1, W3(H$_2$O), and AFGL\,2591) have the largest $R_{\rm GC}$ (the lowest O/H ratio) with the exception of the extreme outer Galaxy source, ranging from 8.4 kpc (AFGL\,2591) to $\sim$10 kpc (W3(H$_2$O)). AFGL\,2591 is associated with the Local arm, while the remaining sources with the Perseus arm (\citealt{reid2019}). $L_{\rm HDO}$ for three out of four sources (NGC\,7538\,IRS1, W3(H$_2$O), and AFGL\,2591) are based on the directly measured HDO 241.6 GHz transition. Based on studies on the radial elemental abundance gradients, the metallicity $Z$ at 10 kpc ranges from 0.5 $Z_{\odot}$ to 1.1 $Z_{\odot}$ depending on the tracers used.  Lower values of $Z$ have been obtained from observations of H\,{\sc ii} regions (e.g., \citealt{rudolph2006}; \citealt{esteban2018}), while the higher values from Cepheids (e.g., \citealt{maciel2019}; \citealt{luck2011}).  While the value of $Z$ at a given $R_{\rm GC}$ is rather uncertain, it is clear that $L_{\rm HDO}$ of the LMC hot cores compares to $L_{\rm HDO}$ of objects located at larger $R_{\rm GC}$ where the oxygen abundance is lower and thus less oxygen is available for chemistry. In fact, the positions of the LMC hot cores N\,105--2\,A and 2\,B fit in the trend seen in the top panel in Fig.~\ref{f:lmcgal} very well for different O/H radial gradients determined in the H\,{\sc ii} region studies (see the Fig.~\ref{f:lmcgal} caption for references), assuming the LMC's value of $12+{\rm log(O/H)}$ of 8.4 (e.g., \citealt{russell1992}).
 
Decreasing $L_{\rm HDO}$ with increasing $R_{\rm GC}$ cannot be attributed solely to decreasing metallicity because $L_{\rm bol}$ shows a similar trend, as demonstrated in the middle panel in Fig.~\ref{f:lmcgal2}. However, a weak metallicity dependence is still present in the $L_{\rm HDO}/L_{\rm bol}$  vs. $Z$ plot (i.e., with the $L_{\rm bol}$ dependence removed; see the lower panel in Fig.~\ref{f:lmcgal2}).  Even though $L_{\rm HDO}$ seems to be largely dependent on source luminosity, metallicity effects also play a role. Based on our data, we are not able to disentangle relative contributions of the bolometric luminosity (temperature) and metallicity (oxygen abundance) effects on $L_{\rm HDO}$.   

We did not find significant differences between Galactic hot cores and the LMC hot cores N\,105--2\,A and 2\,B in terms of HDO;  both $L_{\rm HDO}$ measured toward 2\,A and 2\,B fit in with the $L_{\rm HDO}$  vs. $L_{\rm bol}$ and $L_{\rm HDO}$ vs. $Z$  trends observed toward Galactic hot cores.

\subsection{H$_2$O in the LMC}

\subsubsection{Previous Studies on H$_2$O in the Magellanic YSOs}
\label{s:H2O}

Water has previously been detected in the LMC in the solid phase (ice bands at 3.05 $\mu$m  and 62 $\mu$m; \citealt{vanloon2005}; \citealt{oliveira2006,oliveira2011};  \citealt{shimonishi2008,shimonishi2010,shimonishi2016a}; \citealt{vanloon2010lmc}),  gas-phase (H$_2$O 2$_{12}$--1$_{01}$ and 2$_{21}$--1$_{10}$ transitions at 179.52 $\mu$m and 108.07 $\mu$m; \citealt{oliveira2019}), and as 22 GHz H$_2$O maser emission (interstellar H$_2$O masers in star-forming regions: \citealt{scalise1981,scalise1982}; \citealt{whiteoak1983}; \citealt{whiteoak1986}; \citealt{vanloon2001}; \citealt{lazendic2002}; \citealt{oliveira2006}; \citealt{ellingsen2010}; \citealt{schwarz2012}; \citealt{imai2013}; circumstellar masers in evolved stars: \citealt{vanloon1998,vanloon2001ev,vanloon2012}). 

The water ice studies demonstrated that ice abundances toward massive young stellar objects (YSOs) in the LMC are distinct from those observed toward Galactic YSOs.  In particular, the CO$_2$/H$_2$O column density ratio is two times higher in the LMC compared to the Galaxy \citep{seale2011,gerakines1999}, either due to an overabundance of CO$_2$ or underabundance of H$_2$O.  

\citet{oliveira2009} and \citet{shimonishi2010} argue that the enhanced CO$_2$ production can be the result of the stronger radiation field and/or the higher dust temperature in the LMC; this scenario is supported by laboratory work (e.g., \citealt{dhendecourt1986}) and models of the diffusive grain surface chemistry (e.g., \citealt{ruffle2001}). \citet{shimonishi2016a}'s  `warm ice chemistry' model predicting that high dust temperatures in the LMC suppress the hydrogenation of CO on the grain surface, can reproduce both the enhanced abundance of CO$_2$ and underabundance of CH$_3$OH observed in the LMC. 

However, based on a comparison of the H$_2$O, CO, and CO$_2$ ice column densities between the Galaxy, the LMC, and the Small Magellanic Cloud (SMC), \citet{oliveira2011} concluded that high CO$_2$/H$_2$O column density ratio combined with the relatively unchanged CO--to--CO$_2$ abundances are more consistent with the depletion of H$_2$O rather than an increased production of CO$_2$.  They attribute the depletion of H$_2$O to the combined effects of a lower gas-to-dust ratio and stronger UV radiation field in the LMC: the strong interstellar radiation penetrates deeper into the YSO envelopes as compared with Galactic YSOs, possibly destroying H$_2$O ice (enhancing photodesorption) in less-shielded outer layers, effectively reducing the observed H$_2$O ice column density. The CO$_2$ and H$_2$O ice mixtures that exist deeper in the envelope remain unaffected by the stronger radiation field. 

Far-infrared spectroscopic observations toward massive YSOs in the LMC and SMC with {\it Herschel}/PACS revealed that H$_2$O and OH account for $\sim$10\% of the total line cooling, indicating that the trend of decreasing contribution of H$_2$O and OH cooling from low- to high-luminosity sources observed in the Galaxy (\citealt{karska2014,karska2018}) extends to the massive LMC/SMC YSOs (\citealt{oliveira2019}).

The abundance of 22-GHz H$_2$O masers in the LMC appears to be consistent with that observed in the Galaxy, making them useful signposts of massive star formation  in the LMC in contrast to CH$_3$OH masers which are underabundant (e.g., \citealt{ellingsen2010}).

\begin{figure*}
\centering
\raisebox{4mm}{\includegraphics[width=0.45\textwidth, trim=0 0 0 0, clip]{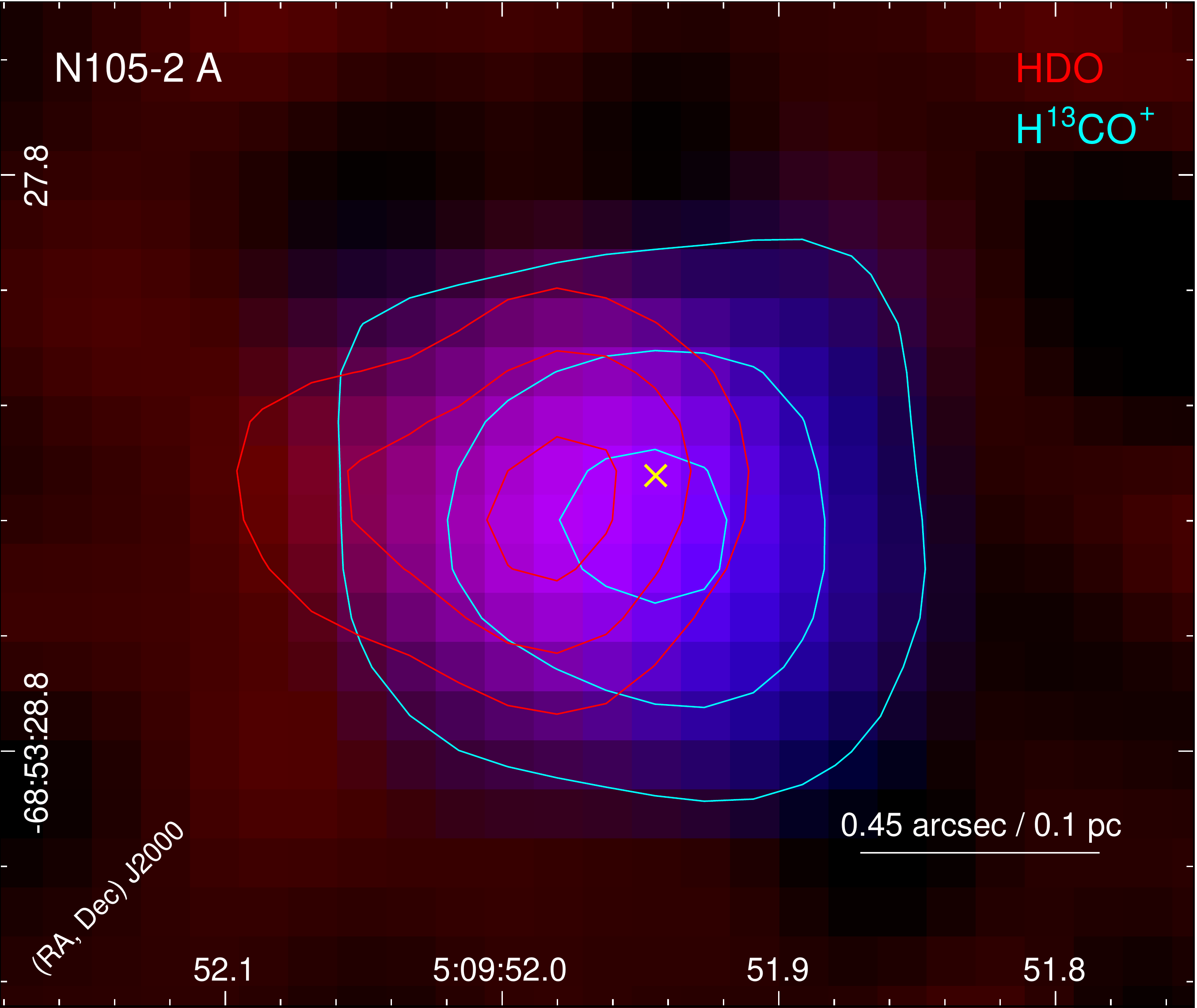}}
\hfill
\includegraphics[width=0.536\textwidth, trim=0 8 0 0, clip]{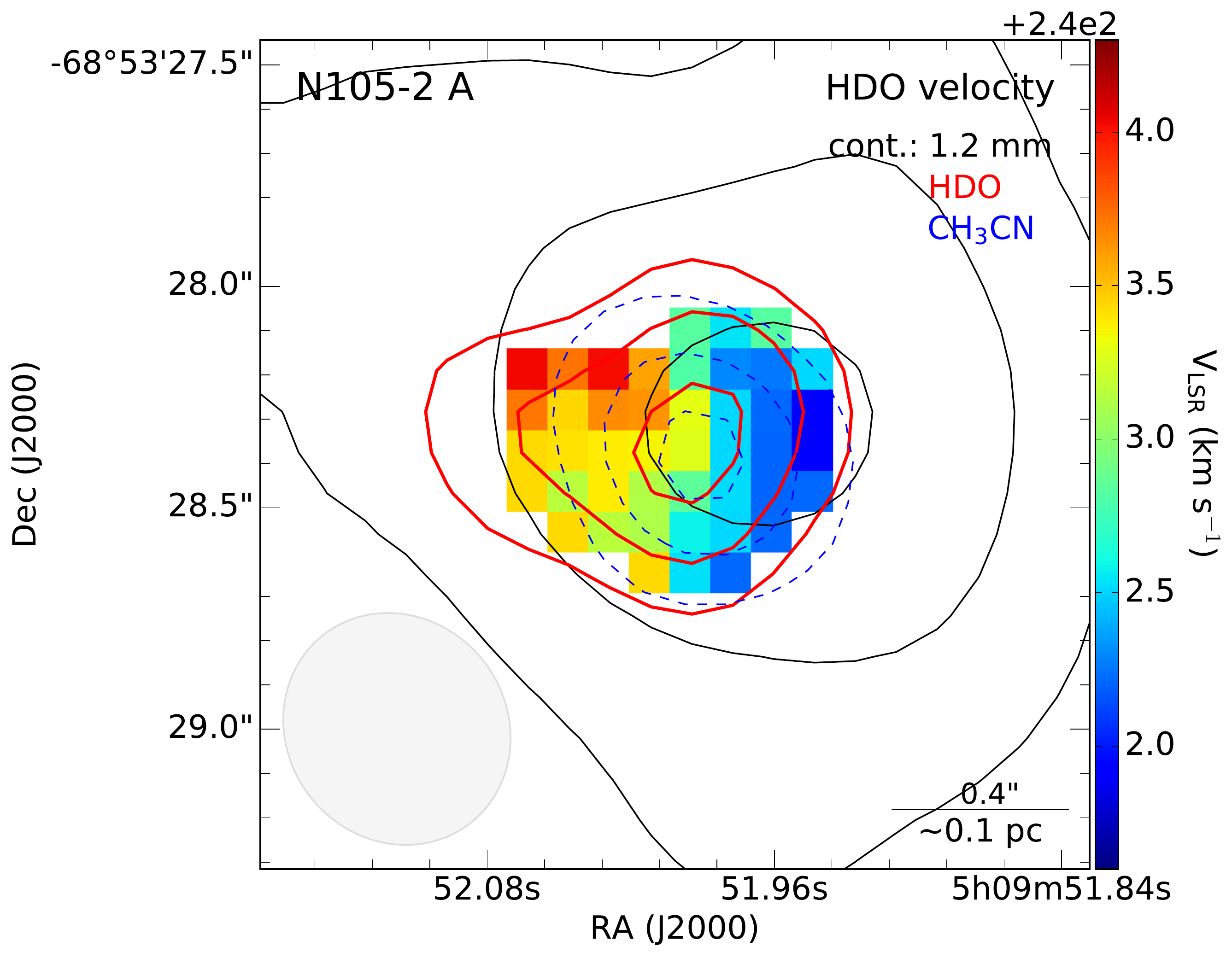}
\caption{{\it Left}: Two-color composite mosaic of N\,105--2\,A combining the integrated intensity images of HDO (red) and H$^{13}$CO$^{+}$ (blue) images. The red and cyan contours correspond to (30, 60, 90)\% of the HDO and H$^{13}$CO$^{+}$ integrated intensity peak of 38.8 and 202.8 mJy beam$^{-1}$ km s$^{-1}$, respectively. The 1.2 mm continuum peak is indicated with an `$\times$' symbol. {\it Right}: The HDO line velocity (moment 1) image of N\,105--2\,A. Red contours are the same as in the left panel. Dashed blue contours correspond to the CH$_3$CN emission with contour levels of (30, 60, 90)\% of the CH$_3$CN integrated intensity peak of 378.6 Jy beam$^{-1}$ km s$^{-1}$.  The 1.2 mm continuum emission contours are shown in black with contour levels of (10, 30, 90)$\sigma$. \label{f:water}}
\end{figure*}

\subsubsection{Estimated H$_2$O Abundance in the LMC Hot Cores N\,105--2\,A and 2\,B}
\label{s:waterabund}

Our observations did not cover any H$_2$O transitions, thus we cannot draw any reliable conclusions regarding the deuterium fractionation (the abundance ratio of deuterated over hydrogenated isotopologues, D/H) of water (HDO/H$_2$O) in the low-metallicity environment; however, since our data did not reveal differences between the Galactic and LMC hot cores N\,105--2\,A and 2\,B based on the analysis of $L_{\rm HDO}$, we made a rough estimate of the H$_2$O column densities and abundances toward 2\,A and 2\,B by assuming that HDO/H$_2$O toward the LMC hot cores is the same as that observed in the Galaxy. 

To date, the deuterium fractionation in the LMC was only determined for DCO$^{+}$ (for star-forming regions N\,113, N\,44\,BC, N\,159\,HW) and DCN (N\,113) on $\sim$10 pc scales (\citealt{chin1996}; \citealt{heikkila1997}). The deuterium fractionation of DCO$^{+}$ ranges from 0.015 to 0.053, while the deuterium fractionation of DCN of 0.043 was found toward N\,113.  These values are similar to those observed toward Galactic dark clouds and pre-stellar cores: 0.01--0.1 (\citealt{ceccarelli2014} and references therein). 

The typical values of water deuteration observed toward Galactic hot cores are of the order of $(2-8)\times10^{-4}$, but they can be as high as $(2-5)\times10^{-3}$ (\citealt{vandishoeck2021} and references therein). For example, HDO/H$_2$O=$(1.2, 0.8, 0.9, 1.6, 3.0)\times10^{-3}$  for (G34.2$+$0.2, W51d, W51e1/e2, Sgr\,B2(N), Orion\,KL) hot cores (\citealt{jacq1990}; \citealt{neill2013}).

We calculated the H$_2$O column densities for 2\,A and 2\,B for the maximum and minimum values in the Galactic HDO/H$_2$O ranges provided above: $5.0\times10^{-3}$ and $2.0\times10^{-4}$, using the HDO column densities for 2\,A and 2\,B provided in Section~\ref{s:results}.
For HDO/H$_2$O of ($5.0\times10^{-3}$, $2.0\times10^{-4}$),  the H$_2$O column densities are 
$N(\rm{H_{2}O})$ $\sim$ ($9.8\times10^{16}$, $2.5\times10^{18}$) cm$^{-2}$ for 2\,A and 
($5.2\times10^{16}$, $1.3\times10^{18}$) cm$^{-2}$  for 2\,B; the abundances with respect to H$_2$ are 
$X(\rm{H_{2}O})$ $\sim$($5.4\times10^{-7}$, $1.4\times10^{-5}$) for 2\,A and 
($1.7\times10^{-7}$, $4.2\times10^{-6}$) for 2\,B.

The typical Galactic hot core H$_2$O abundances range from $5\times10^{-6}$ to $>$10$^{-4}$ \citep{vandishoeck2021}, but a lower value of $1.7\times10^{-6}$ was measured toward IRAS\,16272$-$4837 by \citet{herpin2016}. The metallicity corrected (multiplied by a factor of two, $1/Z_{LMC}$; see \citealt{sewilo2022}) values of $X(\rm{H_{2}O})$ are $\sim$($1.1\times10^{-6}$, $2.8\times10^{-5}$) for 2\,A and  
$\sim$($3.4\times10^{-7}$, $8.4\times10^{-6}$) for 2\,B for the assumed HDO/H$_2$O of ($5.0\times10^{-3}$, $2.0\times10^{-4}$). 

The $X(\rm{H_{2}O})$ range for 2\,A overlaps with the Galactic range for the most part, with the lower end about a factor of two lower than the minimum $X(\rm{H_{2}O})$ measured in the Galactic hot cores. For 2\,B, the $X(\rm{H_{2}O})$ range is shifted toward lower values, down to $X(\rm{H_{2}O})$ of about five times lower than the minimum $X(\rm{H_{2}O})$ measured toward Galactic hot cores. 

We have obtained a similar result for the analysis that only included sources closest in metallicity to that of the LMC and with measured deuterium fractionation of water; these are (W3(H$_2$O), AFGL\,2591, NGC\,7538\,IRS1) with metallicities of $\sim$(0.7, 0.8, 0.7) $Z_{\odot}$ and HDO/H$_2$O of ($(2-6)\times10^{-4}$, $8\times10^{-4}$, $3\times10^{-3}$). The HDO/H$_2$O range for these three sources is basically the same as for the entire population of Galactic disk hot cores.  If we adopt a higher end of the HDO/H$_2$O range of $3\times10^{-3}$ instead of $5\times10^{-3}$ (as above) and scale the estimated LMC value to the average metallicity of W3(H$_2$O), AFGL\,2591, NGC\,7538\,IRS1, the lower end of the $X(\rm{H_{2}O})$ range for 2\,A and 2\,B is 20\% higher, but our conclusions remain the same. 

\subsection{HDO Emission in N105--2: the Spatial Distribution and Velocity Structure} 
\label{s:comp}

The spatial distribution and velocity structure of the HDO emission in N105--2\,A is consistent with HDO being the product of the low-temperature dust grain chemistry. 

In hot cores where the temperatures increase above 100~K, H$_2$O (and HDO) sublimates and becomes an effective destroyer of HCO$^{+}$ (e.g., \citealt{vandishoeck2021} and references therein). This scenario is confirmed in Galactic hot cores where an anti-correlation between H$^{18}_2$O (or CH$_3$OH which is a good proxy for the distribution of water as it desorbs at similar temperature) and H$^{13}$CO$^{+}$ has been observed (e.g., \citealt{jorgensen2013}). We have compared the spatial distribution of HDO and H$^{13}$CO$^{+}$ toward 2\,A and found that HDO and H$^{13}$CO$^{+}$ integrated intensity peaks are separated by $\sim$0$\rlap.{''}$19 ($\sim$0.046 pc or $\sim$9500 AU at 50 kpc; see Fig.~\ref{f:water}), while the peak of the CH$_3$OH emission is coincident with the HDO emission peak. 

In addition to the positional anti-correlation between H$_2$O and HCO$^{+}$, the anti-correlation in velocity is also expected (e.g., \citealt{vandishoeck2021}). For N\,105--2\,A, the (HDO, H$^{13}$CO$^{+}$) velocities are $(242.7\pm0.2, 241.90\pm0.02)$ km s$^{-1}$ (\citealt{sewilo2022}), so there is a small velocity difference of $0.8\pm0.2$ km s$^{-1}$ between HDO and H$^{13}$CO$^{+}$.  The anti-correlation in both the position and velocity between HDO and H$^{13}$CO$^{+}$ toward 2\,A is consistent with the observations of Galactic hot cores and supports the dust grain chemistry origin of HDO in 2\,A. 

The HDO velocity distribution in 2\,A is inconsistent with the shock origin of the HDO emission.  We detected an HDO velocity gradient of $\sim$12 km s$^{-1}$ pc$^{-1}$ that could indicate the presence of an outflow and HDO production in an outflow-driven shock (see Fig.~\ref{f:water}); however, the HDO line is relatively narrow ($4.2\pm0.4$ km s$^{-1}$, $\sim$1 km s$^{-1}$ broader than the H$^{13}$CO$^{+}$ line), making this scenario unlikely. The velocity gradient likely traces the rotation of the core. 

Note that we have not performed the similar analysis for 2\,B because the HDO emission toward this source is much fainter and the results are inconclusive. 

\section{Conclusions} 
\label{s:conclusions}

Based on the analysis of the HDO emission detected toward hot cores N\,105--2\,A and 2\,B in the LMC and a sample of Galactic hot cores covering a range of bolometric luminosities and Galactocentric distances (metallicities), we have found that $L_{\rm HDO}$ measured toward these LMC hot cores follow both the bolometric luminosity and metallicity dependence traced by Galactic sources. Based on our data, we are not able to disentangle the effects of the bolometric luminosity (temperature) and metallicity (oxygen abundance) on $L_{\rm HDO}$, but our results indicate that $L_{\rm bol}$ likely has a larger impact on $L_{\rm HDO}$ than metallicity. 

We have found that if the water deuterium fractionation in the LMC hot cores N\,105--2\,A and 2\,B is within the range observed in the Galactic hot cores, the range of the estimated H$_{2}$O abundances toward 2\,A and 2\,B is shifted toward lower than Galactic values. 

The spatial distribution and velocity structure of the HDO emission in N\,105--2\,A is consistent with HDO being the product of the low-temperature dust grain chemistry.

The astrochemical models of deuterated species predict that HDO is abundant regardless of the extragalactic environment (starburst, cosmic-rays-enhanced environments, low-metallicity, and high-redshift galaxies) and should be detectable with ALMA in many diverse galaxies (\citealt{bayet2010}). Our results for the LMC and the detection of HDO toward the $z$=0.89 absorber against the quasar PKS\,1830$-$211 by \citet{muller2020} are in agreement with these model predictions.  Furthermore, our work demonstrates the utility of HDO as a tracer of H$_2$O chemistry, which is more readily accessible than H$_2$O using ground-based, mm-wave observations.

\begin{acknowledgements}
We thank the anonymous referee for comments that helped us improve the manuscript. The material is based upon work supported by NASA under award number 80GSFC21M0002 (M.S.). A.K. acknowledges support from the First TEAM grant of the Foundation for Polish Science No. POIR.04.04.00-00-5D21/18-00. This article has been supported by the Polish National Agency for Academic Exchange under Grant No. PPI/APM/2018/1/00036/U/001. S.B.C., M.A.C., and E.G.B. were supported by the NASA Planetary Science Division Internal Scientist Funding Program through the Fundamental Laboratory Research work package (FLaRe). 
This research is supported by NSF award 2009624 to U Virginia (R. I.). The National Radio Astronomy Observatory is a facility of the National Science Foundation operated under cooperative agreement by Associated Universities, Inc. This paper makes use of the following ALMA data: ADS/JAO.ALMA\#2019.1.01720.S.  ALMA is a partnership of ESO (representing its member states), NSF (USA) and NINS (Japan), together with NRC (Canada), NSC and ASIAA (Taiwan), and KASI (Republic of Korea), in cooperation with the Republic of Chile. The Joint ALMA Observatory is operated by ESO, AUI/NRAO and NAOJ.  This research made use of APLpy, an open-source plotting package for Python \citep{robitaille2012}.
\end{acknowledgements}

\bibliographystyle{aasjournal}
\bibliography{refs.bib}

\appendix
\counterwithin{figure}{section}

\section{Determination of the HDO Line Luminosity: Data and Methods}
\label{s:appA}

In Table~\ref{t:thedata}, we have compiled the data used for our analysis of the LMC and Milky Way hot cores, both the quantities derived in this paper and the data from literature. The HDO $2_\mathrm{11}-2_\mathrm{12}$ line flux ($F_{\rm HDO\,2_\mathrm{11}-2_\mathrm{12}}$) forms the basis of the analysis, with 11 out of 19 values being directly measured.  $F_{\rm HDO\,2_\mathrm{11}-2_\mathrm{12}}$ for the remaining sources was estimated from the observations of one or two other HDO transitions as described in Section~\ref{s:results}.  We have used $F_{\rm HDO\,2_\mathrm{11}-2_\mathrm{12}}$ to determine the HDO $2_\mathrm{11}-2_\mathrm{12}$ line luminosity ($L_{\rm HDO\,2_\mathrm{11}-2_\mathrm{12}}$). 

The spectral line luminosity ($L$) can be derived based on the line flux (the integrated line intensity) using the standard relation that assumes a Gaussian beam and the Gaussian brightness distribution for the source (e.g., \citealt{wu2005}, their Eq. 2):
\begin{equation}
L =  23.5 \times 10^{-6}  D^2\,  \frac{\pi\,\theta^2_{s}}{4\,{\rm ln}2}\, \frac{\theta^2_{s}+\theta^2_{\rm beam}}{\theta^2_{s}} \int T dv,
\end{equation}
where $D$ is the distance in kpc,  $\theta_{s}$ and $\theta_{\rm beam}$ are the angular sizes in arcseconds of the source and and beam, respectively, and  $\int T dv$ is the line flux in K km~s$^{-1}$.  We calculated $L_{\rm HDO\,2_\mathrm{11}-2_\mathrm{12}}$ from $\int T dv$ = $F_{\rm HDO\,2_\mathrm{11}-2_\mathrm{12}}$, assuming a point source emission and adopting heliocentric distances from the literature. 

In addition to the HDO $2_\mathrm{11}-2_\mathrm{12}$ line fluxes and luminosities, as well as the equatorial and Galactic coordinates, Table~\ref{t:thedata} also lists bolometric luminosities ($L_{\rm bol}$),  distances ($D$), and Galactocentric radii ($R_{\rm GC}$).  All references are provided in the table.

Below, we provide additional information on the analysis of the HDO 1$_\mathrm{10}$--1$_\mathrm{11}$ and 3$_\mathrm{12}$--2$_\mathrm{21}$  data for sources with no observed HDO $2_\mathrm{11}-2_\mathrm{12}$ transition. 

\noindent {\it IRAS\,18089$-$1732, W43\,MM1, W33A, NGC\,7538\,S}:  We used the HDO 1$_\mathrm{10}$--1$_\mathrm{11}$ and 3$_\mathrm{12}$--2$_\mathrm{21}$ data available for these sources to construct the rotational diagram and estimate the HDO $2_\mathrm{11}-2_\mathrm{12}$ line flux (see Section~\ref{s:galsample}). The rotational diagram analysis provided us with the estimate of the HDO rotational temperature ($T_{\rm rot}$) and column density ($N_{\rm HDO}$):  $T_{\rm rot}$ = (82, 78, 110, 90) K and $N_{\rm HDO}$ = $(5.1, 5.0, 6.9, 1.0)\times10^{14}$ cm$^{-2}$ for (IRAS\,18089$-$1732, W43\,MM1, W33A, NGC\,7538\,S).  Since the analysis was based only on two data points (two HDO transitions), we expect the uncertainties to be at least $\sim$50\%. Our result for W33A is fully consistent with \citet{vandertak2006} who analyzed the same HDO data. 

\begin{figure*}[ht!]
\centering
\includegraphics[width=\textwidth]{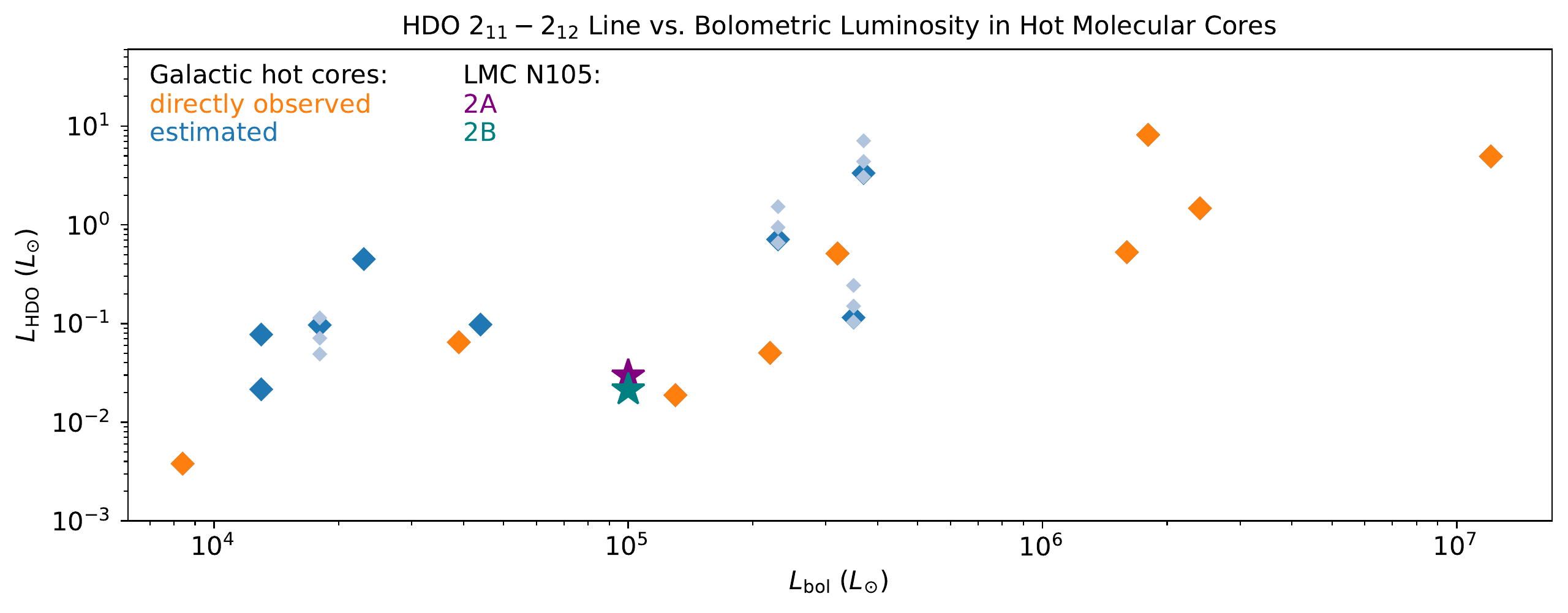} 
\caption{The same as in Fig.~\ref{f:lmcgal} with additional data points shown in light blue for G9.62$+$0.19 ($L_{\rm bol} = 1.8\times10^4$ $L_{\odot}$),  G31.41$+$0.31 ($2.3\times10^5$ $L_{\odot}$), G29.96$-$0.02 ($3.5\times10^5$ $L_{\odot}$), and G10.47$+$0.03A ($3.7\times10^5$ $L_{\odot}$), demonstrating how $L_{\rm HDO}$ for these sources would change if different values of $T_{\rm rot}$ were adopted in the rotational diagram analysis (see Section~\ref{s:galsample}). The data points correspond to  $T_{\rm rot}$ of (from top to bottom):  60, 100, and 200 K. $T_{\rm rot}$ adopted from literature for (G9.62$+$0.19, G31.41$+$0.31, G29.96$-$0.02, G10.47$+$0.03A) is (70, 158, 160, 164) K.  Adopting a different value of $T_{\rm rot}$ does not affect the overall $L_{\rm HDO}$ trend with $L_{\rm bol}$ for the Galactic sample of hot cores. \label{f:lmcgalerr}}
\end{figure*}

\begin{figure*}[ht!]
\centering
\includegraphics[width=\textwidth]{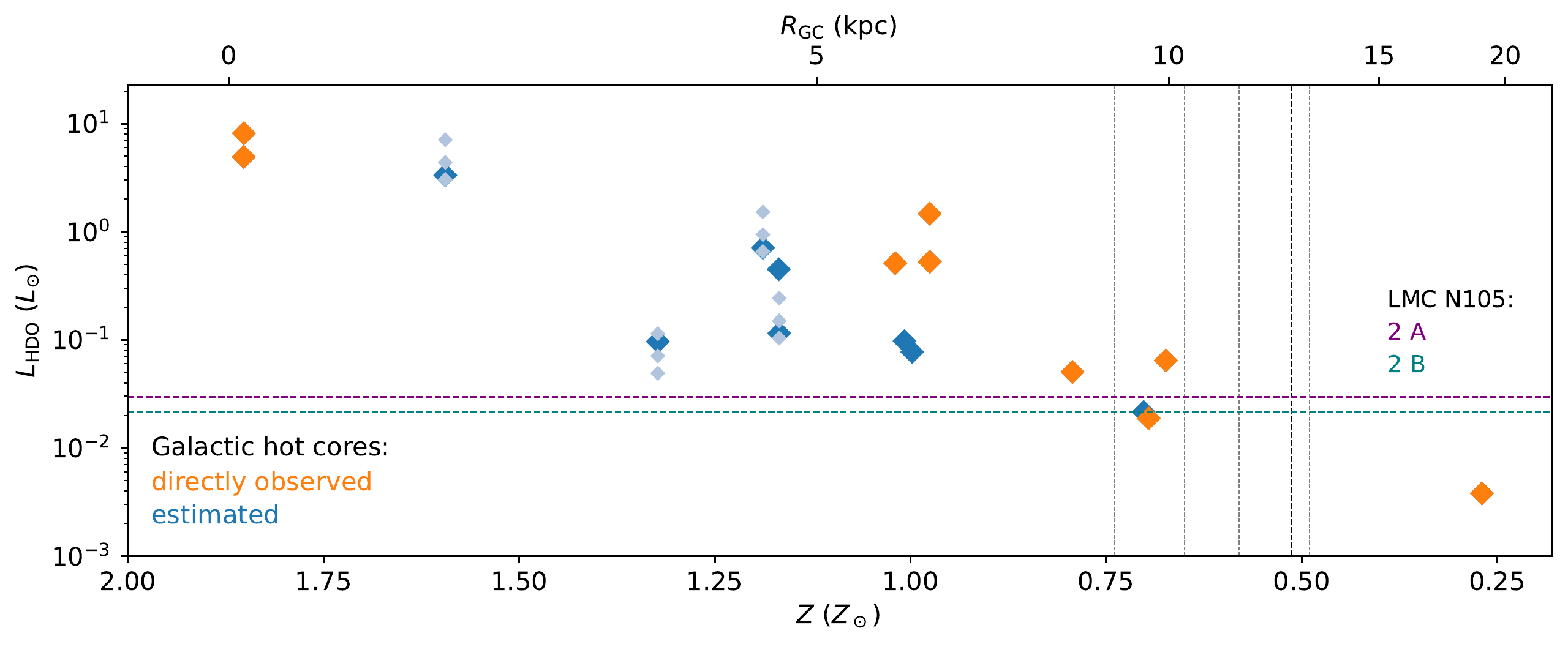}
\includegraphics[width=\textwidth]{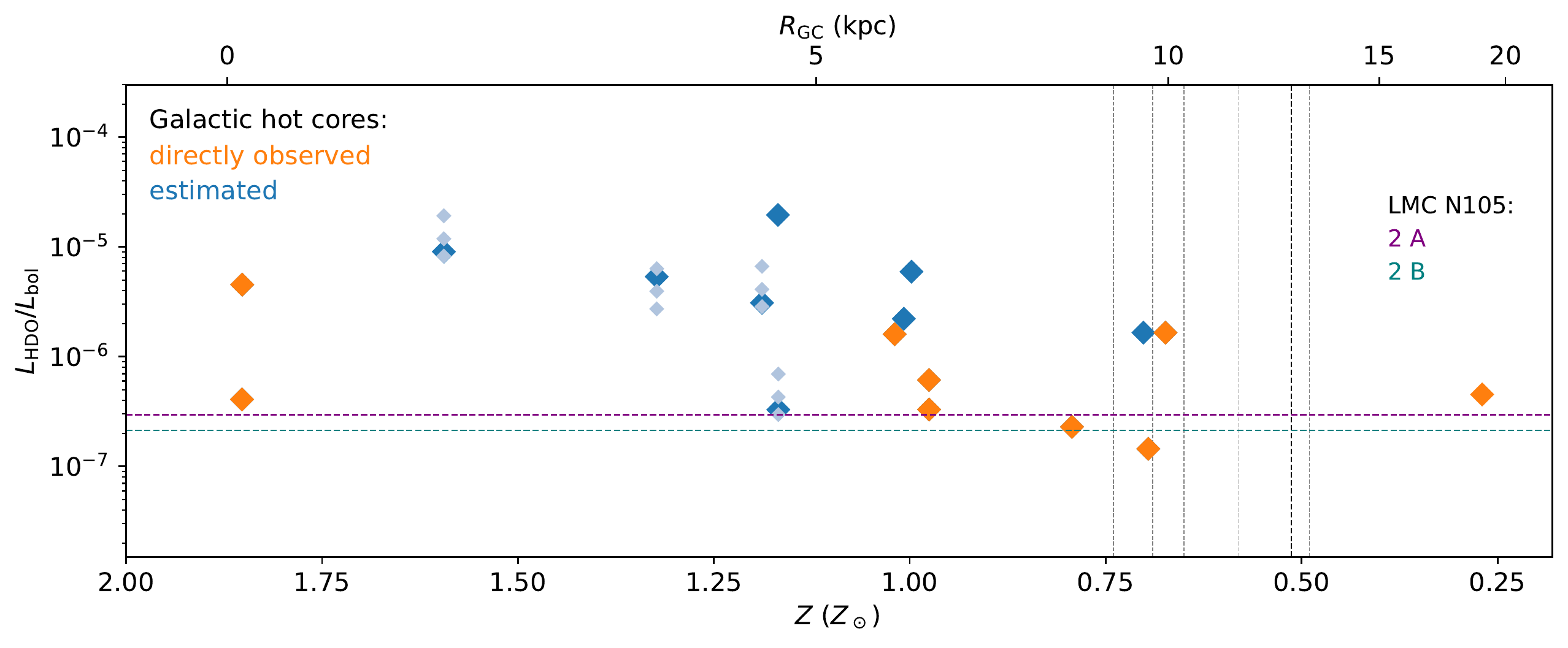}
\caption{The same as in the top and bottom panel in Fig.~\ref{f:lmcgal2} with light blue diamonds as in Fig.~\ref{f:lmcgalerr} for  G10.47$+$0.03A ($Z$$\sim$1.6 $Z_{\odot}$), G9.62$+$0.19 ($\sim$1.3 $Z_{\odot}$), G31.41$+$0.31 ($\sim$1.2 $Z_{\odot}$), and G29.96$-$0.02 ($\sim$1.2 $Z_{\odot}$).  \label{f:lmcgalerr2}}
\end{figure*}

\noindent {\it G9.62$+$0.19, G10.47$+$0.03A, G29.96$-$0.02, G31.41$+$0.31}:   HDO 3$_\mathrm{12}$--2$_\mathrm{21}$ is the only HDO transition available for these sources. As discussed in Section~\ref{s:galsample}, to estimate the HDO 2$_\mathrm{11}$--2$_\mathrm{12}$ line flux, we extrapolated the 3$_\mathrm{12}$--2$_\mathrm{21}$ line flux in the rotational diagram assuming $T_{\rm rot}$ derived in the literature based on CH$_3$CN (\citealt{hofner1996,olmi1996,beltran2011,beltran2005}):  (70, 164, 160, 158) K for (G9.62$+$0.19, G10.47$+$0.03A, G29.96$-$0.02, G31.41$+$0.31). To investigate how adopting a different value of $T_{\rm rot}$ changes $L_{\rm HDO\,2_\mathrm{11}-2_\mathrm{12}}$, we have calculated it for all 4 sources assuming $T_{\rm rot}$ of 60, 100, 200 K. The results are shown in Figs.~\ref{f:lmcgalerr} and \ref{f:lmcgalerr2}.

In Figs.~\ref{f:lmcgalerr} and \ref{f:lmcgalerr2}, we show the same $L_{\rm HDO}$ plots as in Figs.~\ref{f:lmcgal} and \ref{f:lmcgal2}, respectively, with additional data points ($L_{\rm HDO}$ determined for 3 different values of $T_{\rm rot}$) overlaid. The figures show that  
the results for G9.62$+$0.19, G10.47$+$0.03A, G29.96$-$0.02, and G31.41$+$0.31 do not change significantly when different values of temperature are adopted and the conclusions of our work hold. 

\begin{deluxetable*}{lccccccccccccc}
\centering
\rotate
\tablecaption{A Compilation of the data for the LMC N\,105--2\,A and 2\,B hot cores and a sample of Galactic hot cores used in the analysis \label{t:thedata}} 
\tablewidth{0pt}
\tablehead{
\colhead{Hot Core} &
\colhead{RA} &
\colhead{Decl.}&
\colhead{$F_{\rm HDO\,2_\mathrm{11}-2_\mathrm{12}}$} & 
\colhead{$F_{\rm HDO}$ flag\tablenotemark{\footnotesize a}}& 
\colhead{$F_{\rm HDO}$ ref.} & 
\colhead{$L_{\rm HDO\,2_\mathrm{11}-2_\mathrm{12}}$\tablenotemark{\footnotesize b}} & 
\colhead{$L_{\rm bol}$} &
\colhead{$L_{\rm bol}$ ref.} &
\colhead{$D$} &
\colhead{$D$ ref.} &
\colhead{$l$} &
\colhead{$b$} &
\colhead{$R_{\rm GC}$\tablenotemark{\footnotesize c}} \\
\colhead{} &
\colhead{($^{\rm h}$ $^{\rm m}$ $^{\rm s}$)} &
\colhead{($^{\rm \circ}$ $'$ $''$)} &
\colhead{(K km$^{-1}$)} &
\colhead{} &
\colhead{} &
\colhead{(10$^{-2}$ $L_{\odot}$)} &
\colhead{($L_{\odot}$)} &
\colhead{} &
\colhead{(kpc)} &
\colhead{} &
\colhead{($^{\rm \circ}$)} &
\colhead{($^{\rm \circ}$)} &
\colhead{(kpc)}
}
\startdata
\multicolumn{14}{c}{LMC} \\
\hline
N105--2\,A &05:09:51.96&-68:53:28.3&1.8(0.3)&1& this paper &3.0(0.5)& $\sim$$1.0\times10^{5}$ &this paper&50& 16 &279.7526&-34.2520&\nodata\\
N105--2\,B&05:09:52.56&-68:53:28.1&1.3(0.6)&1& this paper &2.2(1.0)& $\sim$$1.0\times10^{5}$ &this paper&50& 16 &279.7523&-34.2511&\nodata\\
\hline
\multicolumn{14}{c}{Milky Way} \\
\hline
IRAS\,18089$-$1732 &18:11:51.5 &$-$17:31:29 &1.24&2& 1 &7.73& $1.3\times10^{4}$ & 9 &2.3& 17 &12.8887&0.4897&6.12\\
NGC\,7538\,S &23:13:44.5&$+$61:26:50&0.23&2& 1 &2.15& $1.3\times10^{4}$ & 10 &2.8& 18 &111.533&0.7568&9.55\\
G9.62$+$0.19 &18:06:15.0&$-$20:31:42&0.31&3&2&9.64& $1.8\times10^{4}$ & 11 &5.15& 19 &9.62&0.19&3.37\\
W43\,MM1 &18:47:47.0&$-$01:54:28&1.26&2& 1 &44.94& $2.3\times10^{4}$ & 9 &5.5& 20 &30.8175&-0.0571&4.58\\
W3\,(H$_2$O)&02:27:03.9&$+$61:52:25&1.2&1&3&6.45& $3.9\times10^{4}$& 12 &2.14& 21 &133.9487&1.0649&9.95\\
W33A &18:14:39.1&$-$17:52:07&1.44&2& 4 &9.75& $4.4\times10^{4}$ & 9 &2.4& 22 &12.9069&-0.2589&6.02\\
NGC\,7538\,IRS1&23:13:45.3&$+$61:28:10&0.7(0.4)&1& 5 &1.88(1.08)& $1.3\times10^{5}$ & 9 &2.65& 23 &111.5422&0.7772&9.63\\
AFGL\,2591 &20:29:24.7&$+$40:11:19&0.394(0.080) &1& 4 & 5.04(1.02)& $2.2\times10^{5}$ & 9 &3.3& 24 &78.8872&0.7085&8.36\\
G31.41$+$0.31 &18:47:34.3&$-$01:12:46&0.97&3& 2 &71.1& $2.3\times10^{5}$ & 9 &7.9& 25 &31.41&0.31&4.42\\
G34.26$+$0.15 &18:53:18.6&$+$01:14:58&12.27(0.05)&1& 6 &51.2(0.2)& $3.2\times10^{5}$ & 9 &3.3& 26 &34.26&0.15&5.91\\
G29.96$-$0.02 &18:46:03.8&$-$02:39:22&0.35&3& 2 &11.49& $3.5\times10^{5}$ & 9 & 5.3 & 27 &29.96&-0.02&4.56\\
G10.47$+$0.03A & 18:08:38.2&$-$19:51:50&3.89&3& 2 &334.35& $3.7\times10^{5}$ & 9 &8.55& 19 &10.47&0.03&1.56\\
W51\,e1/e2 &19:23:43.9&$+$14:30:29&4.7(1.1)&1& 5 &52.8(12.3)& $1.6\times10^{6}$ & 9; 13 &5.41& 28 &49.49&-0.39&6.34\\
Sgr\,B2(N) &17:46:07.9&$+$28:20:12&9.1&1& 7 &815.74& $1.8\times10^{6}$ & 14 &8.34& 29 &0.6773&-0.029&0.099\\
W51\,d &19:23:39.6&$+$14:31:07&13.1(11.0)&1& 5 &147.0(123.5)& $2.4\times10^{6}$ & 15 &5.41& 28 &49.4904&-0.3695&6.34\\
Sgr\,B2(M) &17:46:08.2&$+$28:20:58&5.5&1& 7 &493.03& $1.2\times10^{7}$ & 14 &8.34& 29 &0.6672 &-0.0364&0.097\\
WB89--789\,SMM1 & 06:17:24.07&$+$14:54:42.3&5.05(0.29)&1& 8 &0.38(0.02)& $8.4\times10^{3}$ & 8 &10.7& 30 &195.8219&-0.568&18.86\\
\enddata
\tablerefs{(1) \citet{marseille2010}; (2) \citet{gensheimer1996}; (3) \citep{helmich1996}; (4) \citet{vandertak2006}; (5) \citet{jacq1990}; (6) \citet{coutens2014}; (7) \citet{nummelin2000}; (8) \citet{shimonishi2021}; (9) \citet{vandertak2013}; (1) \citet{wright2012}; (11) \citet{hofner1996}; (12) \citet{ahmadi2018}; (13) \citet{hernandez2014}; (14) \citet{schmiedeke2016}; (15) \citet{rolffs2011}; (16) \citet{pietrzynski2013}; (17) \citet{xu2011}; (18) \citet{sandell2003}; (19) \citet{sanna2014}; (20) \citet{nguyenluong2011}; (21) \citet{navarete2019}; (22) \citet{immer2013}; (23) \citet{moscadelli2008}; (24) \citet{rygl2012}; (25) \citet{churchwell1990}; (26) \citet{kuchar1994}; (27) \citet{zhang2014}; (28) \citet{sato2010}; (29) \citet{reid2014}; (30) \citet{brand2007}}
\tablenotetext{a}{`$F_{\rm HDO}$ flag' indicates whether the HDO $2_\mathrm{11}-2_\mathrm{12}$ line flux ($F_{\rm HDO\,2_\mathrm{11}-2_\mathrm{12}}$) is directly measured from the HDO 2$_\mathrm{11}$--2$_\mathrm{12}$ line observations or estimated based on the observations of other HDO transitions: 1 -- the observed value; the uncertainties are provided when available;  2 -- estimated using the HDO 1$_\mathrm{10}$--1$_\mathrm{11}$ and 3$_\mathrm{12}$--2$_\mathrm{21}$ lines and the rotational diagram; the uncertainties are about 30\%;  3 -- estimated using the HDO 3$_\mathrm{12}$--2$_\mathrm{21}$ line and the rotational diagram, adopting the value of temperature from literature. See Section~ for details.}
\tablenotetext{b}{$L_{\rm HDO\,2_\mathrm{11}-2_\mathrm{12}}$ is the HDO $2_\mathrm{11}-2_\mathrm{12}$ line luminosity calculated using Eq. A1.}
\tablenotetext{c}{$R_{\rm GC}$ is a Galactocentric distance calculated for Galactic hot cores based on their Galactic coordinates ($l$, $b$) and heliocentric distances ($D$; kinematic or parallax), and assuming the distance to the Galactic Center of 8.34 kpc \citep{reid2014}.}
\end{deluxetable*}

\clearpage

\section{Bolometric Luminosity of N\,105--2\,A and 2\,B}
\label{s:lbol}

The multi-wavelength data with high enough spatial resolution to resolve sources N105--2\,A and 2\,B are not available at this time, thus we are not able to determine their individual bolometric luminosities ($L_{\rm bol}$) independently. Instead, we have estimated their combined $L_{\rm bol}$ and inferred their individual contributions based on the highest resolution data. 

To construct the multi-wavelength spectral energy distribution (SED) of the combined sources N105--2\,A and 2\,B (N105--2\,A/2\,B), we have used the seven-band {\it Spitzer} Space Telescope photometric measurements from \citet{gruendl2009} covering 3.6--24 $\mu$m (catalog source 050952.26$-$685327.3; point spread function's full widths at half-maximum, ${\rm FWHMs}\sim1\rlap.{''}7-18''$; \citealt{sage}), five-band {\it Herschel} Space Observatory photometric measurements from \citet{seale2014} covering 100--500 $\mu$m (HSOBMHERICC\,J77.466495-68.891241; ${\rm FWHMs}\sim8\rlap.{''}6-40\rlap.{''}5$; \citealt{heritage}), and a combined ALMA 1.2 mm continuum flux density from Sewi{\l}o et al. (2022). The 1.2 mm flux density has been calculated from the same area used by \citet{gruendl2009} to extract the {\it Spitzer} photometry. N105--2\,A and 2\,B have no counterparts in the near-infrared catalogs such as 2MASS (\citealt{skrutskie2006}; see also \citealt{gruendl2009}) or VISTA VMC \citep{cioni2011}.  

In addition, we have used the {\it Spitzer} InfraRed Spectrograph (IRS) spectrum from \citet{seale2009} to better constrain the SED between 5.2 and 37.9 $\mu$m.  We extracted 11 data points from the IRS spectrum that were selected at wavelengths free of fine-structure emission lines to delineate silicate features and the underlying continuum.  The IRS data points at 20--30 $\mu$m have fluxes $\sim$50\% lower than the MIPS 24 $\mu$m catalog measurement and are likely due to the scaling factors that were applied additionally to match smoothly the spectrum segments taken under different modules across the full wavelength range (\citealt{seale2009}).  We have thus reverted these IRS fluxes to their original values for three affected spectrum segments by removing the corresponding scaling factors, i.e., dividing the fluxes within SL1 (short wavelength, low resolution; 7.6--14.6 $\mu$m), SH (short wavelength, high resolution; 9.9--19.3 $\mu$m), and LH (long wavelength, high resolution; 18.9--36.9 $\mu$m) modules by 1.091, 0.746, and 0.612, respectively.  The resultant IRS fluxes are in good agreement with the MIPS 24 $\mu$m photometric flux measurement from \citet{gruendl2009}.  

The 70 $\mu$m photometry for 050952.26$-$685327.3 is not available in the existing catalogs (SAGE, \citealt{meixner2006}; \citealt{gruendl2009}), therefore we performed an aperture photometry on the SAGE 70 $\mu$m image to estimate the 70 $\mu$m flux of N105--2\,A/2\,B. We used an aperture with a 16$''$ radius, a 39$''$--65$''$ background annulus, and we applied an aperture correction factor of 2.087 (see also \citealt{chen2010}). 

\begin{figure*}[bt!]
\centering
\includegraphics[width=0.6\textwidth]{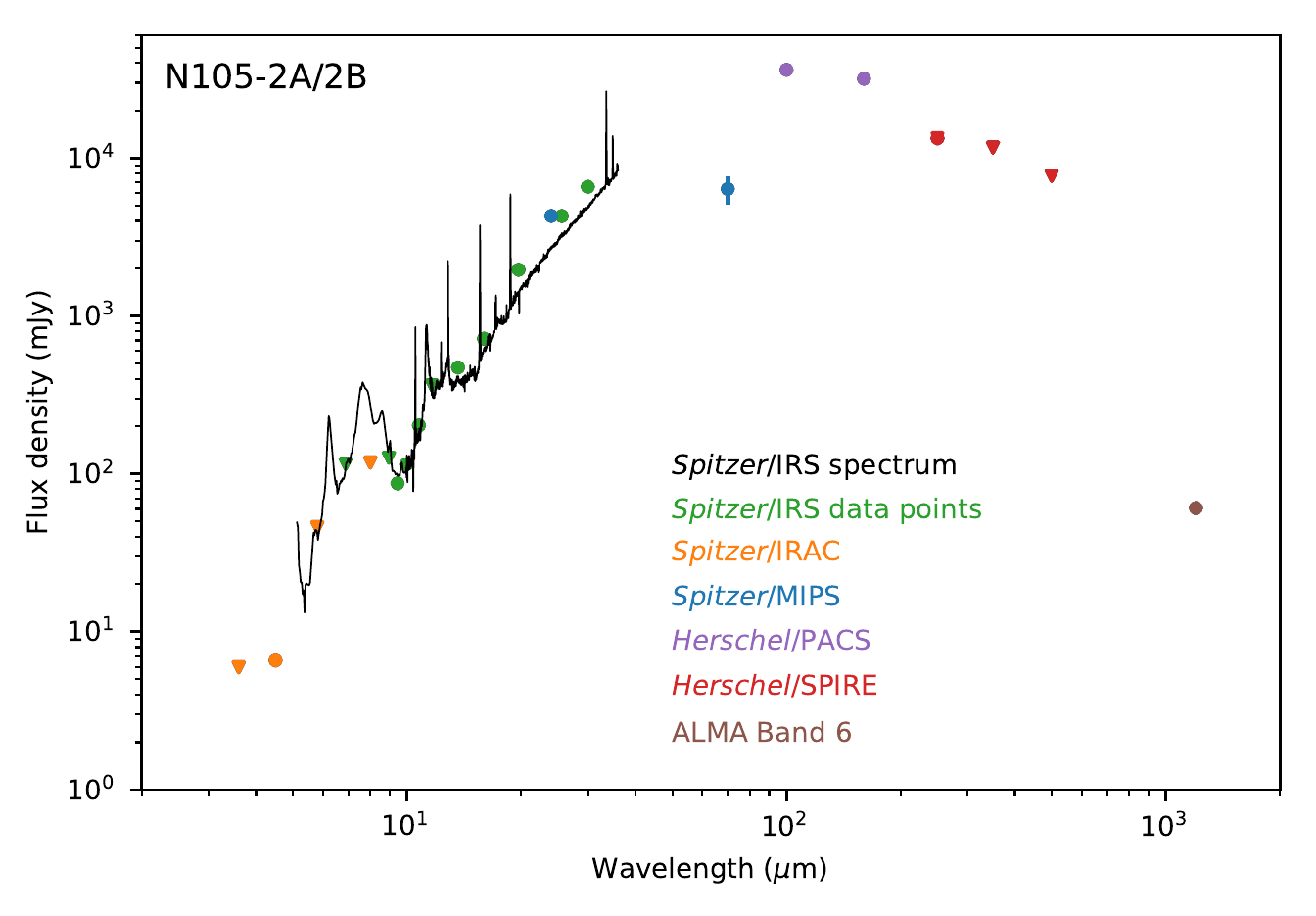}
\caption{The spectral energy distribution (SED) for the combined sources N\,105--2\,A and 2\,B, covering the wavelength range from 3.6 $\mu$m to 1.2 mm.  Filled circles and triangles are valid flux values and flux upper limits, respectively.  The flux error bars are plotted if larger than the data points. \label{f:sed}} 
\end{figure*}

\begin{figure*}[ht!]
\centering
\includegraphics[width=\textwidth]{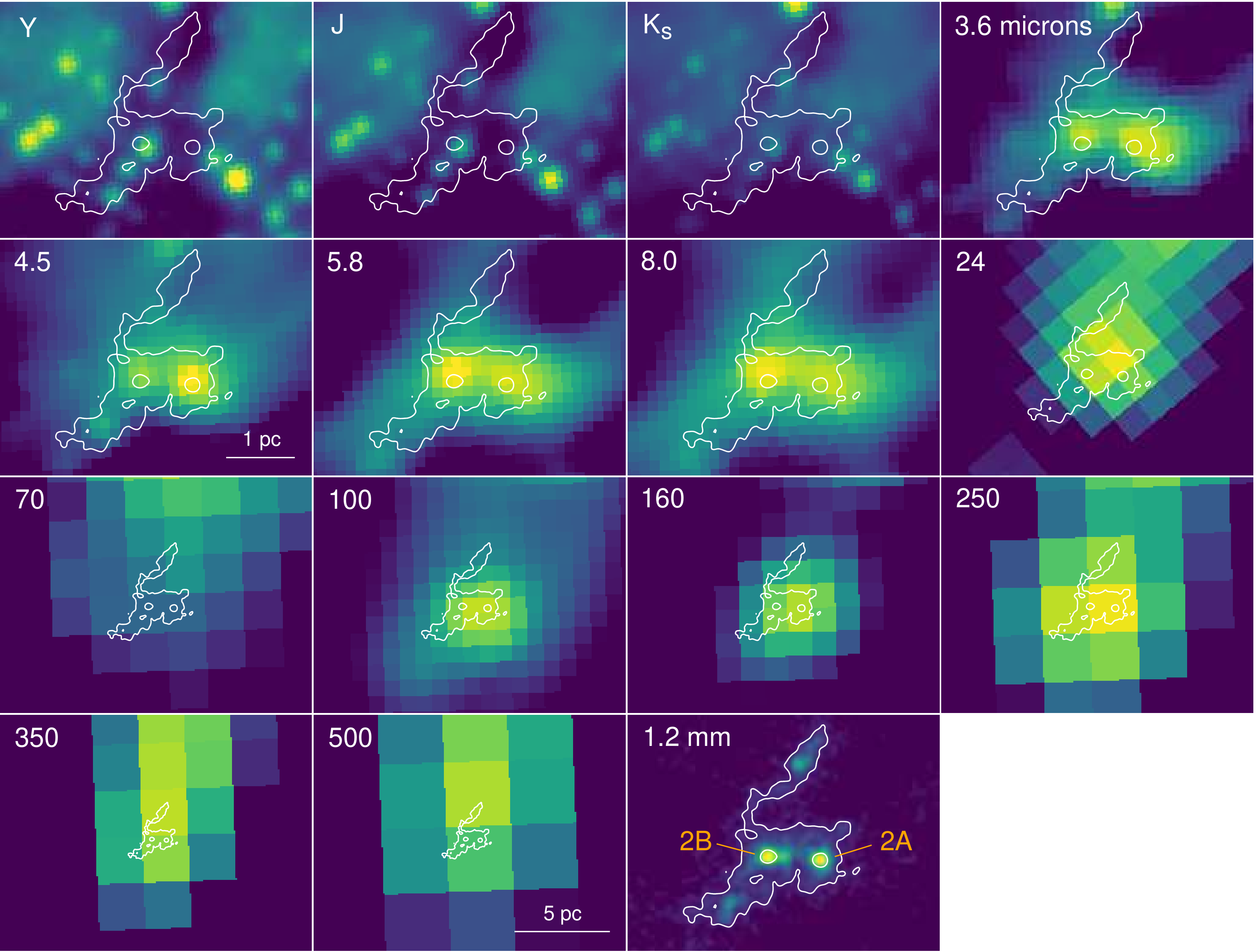}
\caption{Multiwavelength images of N\,105--2\,A and 2\,B ({\it from top left to bottom right}): VISTA VMC $YJK_{s}$ \citep{cioni2011}, {\it Spitzer}/SAGE IRAC 3.6--8.0 $\mu$m and MIPS 24 and 70 $\mu$m (\citealt{meixner2006}; \citealt{sage}), {\it Herschel}/HERITAGE PACS 100 and 160 and SPIRE 250--500 $\mu$m (\citealt{meixner2013}; \citealt{heritage}), and ALMA 1.2 mm \citep{sewilo2022}. The 1.2 mm continuum contours with contour levels of (3, 50) times the image rms of $5.1\times10^{-5}$ mJy beam$^{-1}$ are overlaid on all the images for reference. \label{f:cutouts}} 
\end{figure*}

The SED for N105--2\,A/2\,B and multi-wavelength image cutouts are shown in Fig.~\ref{f:sed} and \ref{f:cutouts}, respectively. 

While 2\,A and 2\,B were extracted as a single {\it Spitzer} source by \citet{gruendl2009}, they are marginally resolved in all {\it Spitzer}/IRAC images (see Fig.~\ref{f:cutouts}). To assess individual flux contributions from N105-2\,A and 2\,B to the combined, unresolved infrared photometric measurements, we carried out aperture photometry of their counterparts on {\it Spitzer}/IRAC images.  As the $\sim$2$''$ separation between these two sources translates to $\sim$1.5 pixels at IRAC's pixel scale, we used 1-pixel radius to estimate their flux ratios.  The 2\,A to 2\,B flux ratios are $\sim$1.4--1.6 at 3.6 and 4.5 $\mu$m and $\sim$1 at 5.8 and 8.0 $\mu$m.  Comparable fluxes at longer {\it Spitzer} wavelengths (5.8 and 8.0 $\mu$m) and the small difference at shorter wavelengths (3.6 and 4.5 $\mu$m) suggest that the two sources are likely to contribute similarly to the unresolved measurements at longer wavelengths.  In addition, 2\,A and 2\,B have the same continuum flux densities at 1.2 mm within the uncertainties. We thus assumed the unresolved fluxes are partitioned equally between 2\,A and 2\,B.

We have estimated $L_{\rm bol}$ in two ways. First, we fitted the SED of N105--2\,A/2\,B with a set of radiative transfer model SEDs for YSOs developed by \citet{robitaille2017} using the \citet{robitaille2007} SED fitting tool. We selected the best-fit model using the procedure outlined in \citet{sewilo2019}; it includes both an envelope and a disk, consistent with the classification of 2\,A and 2\,B as hot cores. Considering the fact that the SED corresponds to two objects, we only use the fitting results to determine luminosity. The 70 $\mu$m flux has a large uncertainty that can only be improved with higher-resolution observations.  It is difficult to judge whether the 70 $\mu$m flux is a lower or an upper limit (see Fig.~\ref{f:cutouts}) and hence the data point carries little weight in the fitting.  We have obtained $L_{\rm bol}$ of $\sim$10$^5$ $L_{\odot}$ for each N105--2\,A and N105--2\,B.

To estimate $L_{\rm bol}$, we also used the trapezoidal method to sum up the area under the SED resulting in $L_{\rm bol}$ of $2.4\times10^5 L_{\odot}$, consistent with the SED fitting results. In this method, we excluded the 70 $\mu$m flux and treated all the remaining fluxes as valid data points. 

\end{document}